\pdfoutput=1

\documentclass[11pt,twoside,a4paper,cmspaper,final,collab]{cms-tdr}
\def\svnVersion{455003P}\def\cmsCernNoTag{CERN-EP-2017-184}\def\cmsCernDate{\today}\def\cmsMessage{Published in Physical Review D as \href{http://dx.doi.org/10.1103/PhysRevD.97.072006}{\doi{10.1103/PhysRevD.97.072006.}}}

\begin{document}\cmsNoteHeader{B2G-17-001}

\hyphenation{had-ron-i-za-tion}
\hyphenation{cal-or-i-me-ter}
\hyphenation{de-vices}
\RCS$HeadURL: svn+ssh://svn.cern.ch/reps/tdr2/papers/B2G-17-001/trunk/B2G-17-001.tex $
\RCS$Id: B2G-17-001.tex 454976 2018-04-11 15:48:38Z dschafer $
\newlength\cmsFigWidth
\ifthenelse{\boolean{cms@external}}{\setlength\cmsFigWidth{0.38\textwidth}}{\setlength\cmsFigWidth{0.45\textwidth}}
\ifthenelse{\boolean{cms@external}}{\providecommand{\cmsLeft}{top\xspace}}{\providecommand{\cmsLeft}{left\xspace}}
\ifthenelse{\boolean{cms@external}}{\providecommand{\cmsRight}{bottom\xspace}}{\providecommand{\cmsRight}{right\xspace}}
\providecommand{\NA}{---}
\newcommand{\x}{\ensuremath{\phantom{0}}}
\newcommand{\y}{\ensuremath{\phantom{.}}}
\newcommand{\PV}{\ensuremath{\mathrm{V}}}
\newcommand{\WW}{\ensuremath{\PW\PW}\xspace}
\ifthenelse{\boolean{cms@external}}{\newcommand{\WZ}{\ensuremath{\PW\!\PZ}\xspace}}{\newcommand{\WZ}{\ensuremath{\PW\PZ}\xspace}}
\ifthenelse{\boolean{cms@external}}{\newcommand{\ZZ}{\ensuremath{\PZ\!\PZ}\xspace}}{\newcommand{\ZZ}{\ensuremath{\PZ\PZ}\xspace}}
\ifthenelse{\boolean{cms@external}}{\newcommand{\VV}{\ensuremath{\PV\!\PV}\xspace}}{\newcommand{\VV}{\ensuremath{\PV\PV}\xspace}}
\newcommand{\SRLOW}{65\xspace}
\newcommand{\SRMIDDLE}{85\xspace}
\newcommand{\SRHIGH}{105\xspace}
\newcommand{\SFWTAGHPWPT} {$0.99 \pm 0.1\stat \pm 0.04\syst$}
\newcommand{\SFWTAGLPWPT} {$1.03 \pm 0.2\stat \pm 0.11\syst$}
\newcommand{\LUMIUNCERT} {2.5\%\xspace}
\newcommand{\Wo}{\PW\xspace}
\newcommand{\Zo}{\PZ\xspace}
\newcommand{\qo}{\ensuremath{\PQq}\xspace}
\newcommand{\mVV}{\ensuremath{m_{\VV}}\xspace}
\newcommand{\mqV}{\ensuremath{m_{\qo\PV}}\xspace}
\newcommand{\mjj}{\ensuremath{m_\mathrm{jj}}}
\newcommand{\mJ}{\ensuremath{m_{\text{jet}}}}
\newcommand{\nsubj}{\ensuremath{\tau_{21}}}
\newcommand{\ktilde}{\ensuremath{\tilde{k}}}
\newcommand{\BulkG}{\ensuremath{\PXXG_{\text{bulk}}}\xspace}
\newcommand{\INTLUMI}     {35.9\fbinv}

\cmsNoteHeader{B2G-17-001}
\title{Search for massive resonances decaying into \texorpdfstring{$\WW$, $\WZ$, $\ZZ$, $\PQq\PW$, and $\PQq\PZ$}{WW, WZ,ZZ, qW, and qZ} with dijet final states at \texorpdfstring{$\sqrt{s} = 13\TeV$}{sqrt(s) = 13 TeV}}

\date{\today}

\abstract{
Results are presented from a search in the dijet final state for new massive narrow resonances decaying to pairs of $\PW$ and $\PZ$ bosons or to a $\PW\!/\PZ$ boson and a quark. Results are based on data recorded in proton-proton collisions at $\sqrt{s} = 13\TeV$ with the CMS detector at the CERN LHC. The data correspond to an integrated luminosity of 35.9\fbinv. The mass range investigated extends upwards from 1.2\TeV. No excess is observed above the estimated standard model background and limits are set at 95\% confidence level on cross sections, which are interpreted in terms of various models that predict gravitons, heavy spin-1 bosons, and excited quarks. In a heavy vector triplet model, \PWpr{} and \PZpr{} resonances, with masses below 3.2 and 2.7\TeV, respectively, and spin-1 resonances with degenerate masses below 3.8\TeV are excluded at 95\% confidence level. In the case of a singlet \PWpr{} resonance masses between 3.3 and 3.6\TeV can be excluded additionally. Similarly, excited quark resonances, $\PQq^*$, decaying to $\PQq\PW$ and $\PQq\PZ$ with masses less than 5.0 and 4.7\TeV, respectively, are excluded. In a narrow-width bulk graviton model, upper limits are set on cross sections ranging from 0.6\unit{fb} for high resonance masses above 3.6\TeV, to 36.0\unit{fb} for low resonance masses of 1.3\TeV.
}

\hypersetup{%
pdfauthor={CMS Collaboration},%
pdftitle={Search for massive resonances decaying into WW, WZ, ZZ, qW, and qZ channels in dijet final states at sqrt(s) = 13 TeV},%
pdfsubject={CMS},%
pdfkeywords={CMS, physics}}

\maketitle

\section{Introduction}
\label{sec:introduction}

The standard model (SM) of particle physics describes with high accuracy a multitude of experimental and observational data.
Nevertheless, the SM does not accommodate phenomena such as gravity or dark matter and dark energy inferred from cosmological observations, prompting theoretical work on its extensions. Theories that address these shortcomings commonly predict new particles, which can potentially be observed at the CERN LHC. Models in which these new particles decay to $\VV$ or $\PQq\PV $, where $\PV$ denotes either a $\PW$ or a $\PZ$ boson, are considered in this work.
Searches for diboson resonances have previously been performed in many different final states, placing lower limits above the TeV scale on the masses of these resonances~\cite{Khachatryan:2014hpa,Khachatryan:2014gha,ATLASwprimeWZPAS,Khachatryan:2014xja,Khachatryan:2015ywa,Sirunyan:2017wto,Khachatryan:2016cfx,Aad:2014xka,Aad:2015ufa,Aad:2015yza,Aad:2015owa,Khachatryan:2015bma,Dias:2015mhm,Khachatryan:2016yji,Khachatryan:2016cfa,Aaboud:2016okv,Sirunyan:2016cao,ATLASVH13,Aaboud:2017ahz,Aaboud:2017eta}.
In addition, we consider excited quarks $\PQq^*$~\cite{Bauer1987,PhysRevD.42.815} that decay into a quark and either a $\PW$ or a $\PZ$ boson.
Results from previous searches for such signals include limits placed on the production of $\PQq^*$ at the LHC in the dijet~\cite{PhysRevD.87.114015,Aad2013,Sirunyan:2016iap,Aaboud:2017yvp,Harris:2011bh}, $\gamma$+jet~\cite{Aad:2013cva,Aad:2015ywd,Khachatryan:2014aka}, $\PQq\PW$, and $\PQq\PZ$~\cite{Chatrchyan:2012ypy,Chatrchyan:2012tw} channels.

This paper presents a search for narrow resonances with $\PW$ or $\PZ$ bosons decaying hadronically at resonance masses larger than 1.2\TeV.
The results are applicable to models predicting narrow resonances and are compared to several benchmark models. The analysis is based on proton-proton collision data at $\sqrt{s} = 13\TeV$ collected by the CMS experiment at the LHC during 2016, corresponding to an integrated luminosity of \INTLUMI. We consider final states produced when a $\VV$ boson pair decays into four quarks or $\PQq\PV$ decays into three quarks, and each boson is reconstructed as a single jet, resulting in events with two reconstructed jets (dijet channel).

The analysis exploits the large branching fraction of vector boson decays to quark final states. Due to the large masses of the studied resonances, the boson decay products are highly collimated and reconstructed as single, large-radius jets. Jet substructure techniques, referred to as jet \emph{$\PV$ tagging} ~\cite{Khachatryan:2014vla,CMS-PAS-JME-14-002,JME16003} in the following, are employed to suppress the SM backgrounds, which largely arise from the hadronization of single quarks and gluons.  As in Ref.~\cite{Sirunyan:2016cao} at $\sqrt{s} = 13\TeV$, and Ref.~\cite{Khachatryan:2014hpa} at $\sqrt{s} = 8\TeV$, the analysis presented here searches for a local enhancement in the diboson or quark-boson invariant mass spectrum reconstructed from the two jets with the largest transverse momenta in the event. Compared to the previous measurement ~\cite{Sirunyan:2016cao}, this analysis not only profits from an increase in integrated luminosity of more than a factor of 13 but also uses improved substructure variables.

\section{The CMS detector}
\label{sec:cmsdetector}

The central feature of the CMS apparatus is a superconducting solenoid of 6\unit{m} internal diameter, providing a magnetic field of 3.8\unit{T}. Contained within the superconducting solenoid volume are a silicon pixel and strip tracker, a lead tungstate crystal electromagnetic calorimeter (ECAL), and a brass and scintillator hadron calorimeter (HCAL), each composed of a barrel and two endcap sections. Muons are detected in gas-ionization chambers embedded in the steel flux-return yoke outside the solenoid. Extensive forward calorimetry complements the coverage provided by the barrel and endcap detectors.

The particle-flow (PF) event algorithm reconstructs and identifies each individual particle with an optimized combination of information from the various elements of the CMS detector~\cite{Sirunyan:2017ulk}. The energy of photons is directly obtained from the ECAL measurement, corrected for zero-suppression effects as described in Ref.~\cite{Sirunyan:2017ulk}. The energy of electrons is determined from a combination of the electron momentum at the primary interaction vertex as determined by the tracker, the energy of the corresponding ECAL cluster, and the energy sum of all bremsstrahlung photons spatially compatible with originating from the electron track. The energy of muons is obtained from the curvature of the corresponding track. The energy of charged hadrons is determined from a combination of their momentum measured in the tracker and the matching ECAL and HCAL energy deposits, corrected for zero-suppression effects and for the response function of the calorimeters to hadronic showers. Finally, the energy of neutral hadrons is obtained from the corresponding corrected ECAL and HCAL energy.

A more detailed description of the CMS detector, together with a definition of the coordinate system used and the relevant kinematic variables, can be found in Ref.~\cite{Chatrchyan:2008aa}.

\section{Simulated samples}
\label{sec:simulatedsamples}

Signal samples were generated for the following benchmark models for resonant diboson production: the bulk scenario ($\rm G_{bulk}$)~\cite{Agashe:2007zd, Fitzpatrick:2007qr, Antipin:2007pi} of the Randall-Sundrum (RS) model of  warped extra dimensions~\cite{Randall:1999ee,Randall:1999vf}, as well as vector singlets (\PWpr or \PZpr)~\cite{Pappadopulo:2014qza}, and excited quark resonances $\PQq^*$~\cite{Bauer1987,PhysRevD.42.815} decaying to $\PQq\PW$ or $\PQq\PZ$.

The bulk RS model is described by two free parameters: the mass of the first Kaluza-Klein~(KK) excitation of a spin-2 boson (the KK bulk graviton) and
the ratio $\tilde{k} \equiv k/\overline{M}_{\rm Pl}$, where $k$ is the unknown curvature scale of the extra dimension and $\overline{M}_{\rm Pl} \equiv M_{\rm Pl}/\sqrt{\smash[b]{8\pi}}$ is the reduced Planck mass.
The samples used in this study have $\ktilde = 0.5$~\cite{Oliveira:2014kla}.

The heavy vector triplet (HVT) model generically subsumes a large number of models predicting additional gauge bosons, such as composite Higgs~\cite{Bellazzini:2014yua,CHM2,Composite2,Greco:2014aza,Lane:2016kvg} and little Higgs~\cite{Schmaltz:2005ky,ArkaniHamed:2002qy} models. The specific models that predict \PWpr~\cite{Grojean:2011vu}, \PZpr ~\cite{Salvioni:2009mt}, or \PWpr and \PZpr ~\cite{Pappadopulo:2014qza} resonances are expressed in terms of a few parameters: the strength of the couplings to fermions, $c_{\rm F}$, couplings to the Higgs and longitudinally
polarized SM vector bosons, $c_{\rm H}$, and the interaction strength $g_{\rm V}$ of the new vector boson. Samples were simulated in HVT model B with $g_{\rm V}=3$, $c_{\rm H}=-0.976243$, and $c_{\rm F}=1.02433$. For these model parameters, the new resonances are narrow and have large branching fractions to boson pairs, while the fermionic couplings are suppressed. This scenario is the most representative of a composite Higgs model.
In the HVT and bulk graviton models, the vector bosons are produced with a longitudinal polarization in more than 99\% of the cases, resulting in a ${\sim} 24\%$ higher acceptance per boson than for models producing transversally polarized vector bosons~\cite{Khachatryan:2014vla,Sirunyan:2016cao}. In the case of excited quarks, unpolarized bosons are simulated with the compositeness scale $\Lambda$ equal to the resonance mass.

We restrict the analysis to scenarios where the natural width of the resonance is sufficiently small to be neglected when compared to the detector resolution.
This makes our modeling of the detector effects on the signal shape independent of the actual model used for generating the events. All simulated samples are produced with a relative resonance width of 0.1\%, in order to be firmly in the regime where the natural width is much smaller than the detector resolution.
The Monte Carlo (MC) simulated samples  of signal events for HVT and bulk graviton production are generated with the leading-order (LO) mode of \MGvATNLO{} v5.2.2.2~\cite{Alwall:2014hca}.
The $\PQq^*$ to $\PQq\PW$ and $\PQq\PZ$ processes are generated to LO using \PYTHIA{} version 8.212~\cite{Sjostrand:2007gs}.

Simulated samples of the SM background processes are used to optimize the analysis.
The production of quantum chromodynamics (QCD) multijet events as well as of SM W+jets and Z+jets processes is simulated to leading order with \MGvATNLO~\cite{Alwall:2014hca,Alwall:2007fs}.
The NNPDF 3.0~\cite{Ball:2014uwa} parton distribution functions (PDF) are used for all simulated samples.
All samples are processed through a \GEANTfour-based~\cite{Agostinelli:2002hh} simulation of the CMS detector.
To simulate the effect of additional proton-proton collisions within the same or adjacent bunch crossings (pileup), additional inelastic events are generated using \PYTHIA and superimposed on the hard-scattering events. The MC simulated events are weighted to reproduce the distribution of the number of pileup interactions observed in data, with an average of 21 reconstructed collisions per beam crossing.

\section{Reconstruction and selection of events}
\label{sec:eventreconstruction}

\subsection{Jet reconstruction}

Hadronic jets are constructed from the four-momenta of the PF candidates in an event,
using the \textsc{FastJet} software package~\cite{Cacciari:2011ma}.
Jets used for identifying the hadronically decaying $\PW$ and \Zo bosons
are clustered using the anti-\kt algorithm~\cite{Cacciari:2008gp}
with a distance parameter $R = 0.8$ (AK8 jets). Charged particles identified as originating from pileup vertices are excluded.
A correction based on the area of the jet, projected on the front face of
the calorimeter, is used to take into account the extra energy
clustered in jets due to neutral particles coming from pileup~\cite{Cacciari:2011ma}.
The jet momentum is determined as the vectorial sum of all particle momenta in this jet.
The jet energy resolution amounts typically to 8\% at 100\GeV, and 4\% at 1\TeV~\cite{Khachatryan:2016kdb}.
Additional quality criteria are applied to the jets in order to remove spurious
jetlike features originating from isolated noise patterns in the calorimeters or the
tracker. The efficiency of these jet quality requirements for signal events is above 99\%.
All jets must have transverse momentum $\pt > 200\GeV$ and pseudorapidity $\abs{\eta}<2.5$ in order to be considered in the subsequent steps of the analysis, ensuring that sufficient boson decay products are contained in the jets to allow $\PV$ tagging.

In order to mitigate the effect of pileup on the two jet observables used in the identification of hadronic $\PW$ and $\PZ$ decays (see below for details), we take advantage of pileup per particle identification (PUPPI)~\cite{Bertolini:2014,JME16003}, obviating area-based pileup corrections.
This method uses local shape information such as the local shape of charged pileup, event pileup properties, and tracking information together in order to rescale the four-momentum of neutral PF candidates according to the degree to which the particle is compatible with an origin outside of the primary interaction.
Using the PUPPI method for the calculation of these jet observables leads to a greater robustness against additional hadronic activity.

\subsection{\texorpdfstring{$\PW\to \PQq\PAQq'$ and $\PZ\to \PQq\PAQq$}{W to q anti-q' and Z to q anti-q} identification using jet substructure}
\label{subsec:Vhadr}

The variables used to identify $\PW$ and $\PZ$ jet candidates are reconstructed from AK8 jets with PUPPI pileup mitigation applied, decreasing the dependence on pileup of these variables as shown in Ref.~\cite{JME16003}.
In order to discriminate against multijet backgrounds, we exploit both the reconstructed jet mass, which is required to be close to the $\PW$ or \Zo boson mass, and the two-prong jet substructure produced by the particle cascades of two high-\pt quarks merging into one jet~\cite{Khachatryan:2014vla}. Jets that are identified as coming from the merged decay products of a single $\PV$ boson are hereafter referred to as \emph{$\PV$ jets}.

As the first step in exploring potential substructure, the jet constituents are subjected to a jet grooming algorithm that eliminates soft, large-angle QCD radiation and thereby improves the resolution in the $\PV$ jet mass, lowers the mass of jets initiated by single quarks or gluons coming from multijet background, and reduces the residual effect of pileup~\cite{Chatrchyan:2013vbb,CMS-PAS-JME-14-002}.
In this paper, we use a modified mass-drop algorithm~\cite{jetmass_nnll1,Butterworth:2008iy}, known as the \emph{soft-drop} algorithm~\cite{Larkoski:2014wba}. This method
accomplishes jet grooming in a way that ensures the absence of nonglobal logarithmic terms in the jet mass~\cite{jetmass_nnll1,jetmass_nnll2} in contrast to the \emph{jet-pruning} algorithm~\cite{jetpruning1,Ellis:2009me} used in the previous version of this analysis~\cite{Sirunyan:2016cao}, while providing similar discrimination power~\cite{JME16003}.
The soft-drop algorithm starts from a Cambridge-Aachen (CA)~\cite{Catani:1993hr,Wobisch:1998wt} jet $j$ clustered from the constituents of the original AK8 jet. It breaks the jet into two subjets. If the subjets pass the soft-drop condition defined in Ref.~\cite{Larkoski:2014wba}, $j$ is considered as the final soft-drop jet, otherwise the procedure is iteratively continued on the subjets using the harder of the two subjets as new $j$ and dropping the other subjet until the soft-drop condition is met.
This algorithm is used for the offline analysis while the \textit{jet-trimming} algorithm~\cite{Krohn:2009th} is used at trigger level.
Jet trimming reclusters each AK8 jet starting from all its original constituents using the $\kt$ algorithm~\cite{Catani:1993hr,Ellis:1993tq} to create subjets with a size parameter $R_{\text{sub}}$ set to 0.2, discarding any subjet with $\pt^{\text{subjet}}/\pt^{\text{jet}}< 0.03$.
The algorithm is used at the trigger level, since it can be tuned such that it is slightly more inclusive than the more powerful pruning and soft-drop algorithms, which are used in the subsequent offline analysis where their performance can therefore be studied in detail.

The soft-drop jet mass $\mJ$ used in the analysis is computed from the sum of the four-momenta of the constituents passing the grooming algorithm and weighted according to the PUPPI algorithm; it is then corrected by a factor derived in simulated $\PW$ boson samples to ensure a $\pt$- and $\eta$-independent jet mass distribution centered on the nominal $\PV$ mass. The corrections are factorized into two contributions, one of which is applied to data and simulation and represents a global calibration factor, and another factor which is only applied to simulation that corrects for discrepancies between data and simulation.
The jet is considered as a $\PV$ jet candidate if \mJ{} falls in the range $\SRLOW < \mJ < \SRHIGH\GeV$, which we define as the signal jet mass window.

We additionally employ the so-called \textit{N-subjettiness}~\cite{Thaler:2010tr,JME16003} variable, $\nsubj = \tau_2 / \tau_1$, to reject background jets arising from the hadronization of single quarks or gluons.
Jets coming from hadronic $\PW$ or \Zo decays in signal events are characterized by lower values of $\nsubj$ compared to jets from the SM backgrounds.

\subsection{Trigger and primary vertex selection}

\begin{figure*}[tb]
\centering
\includegraphics[width = \cmsFigWidth]{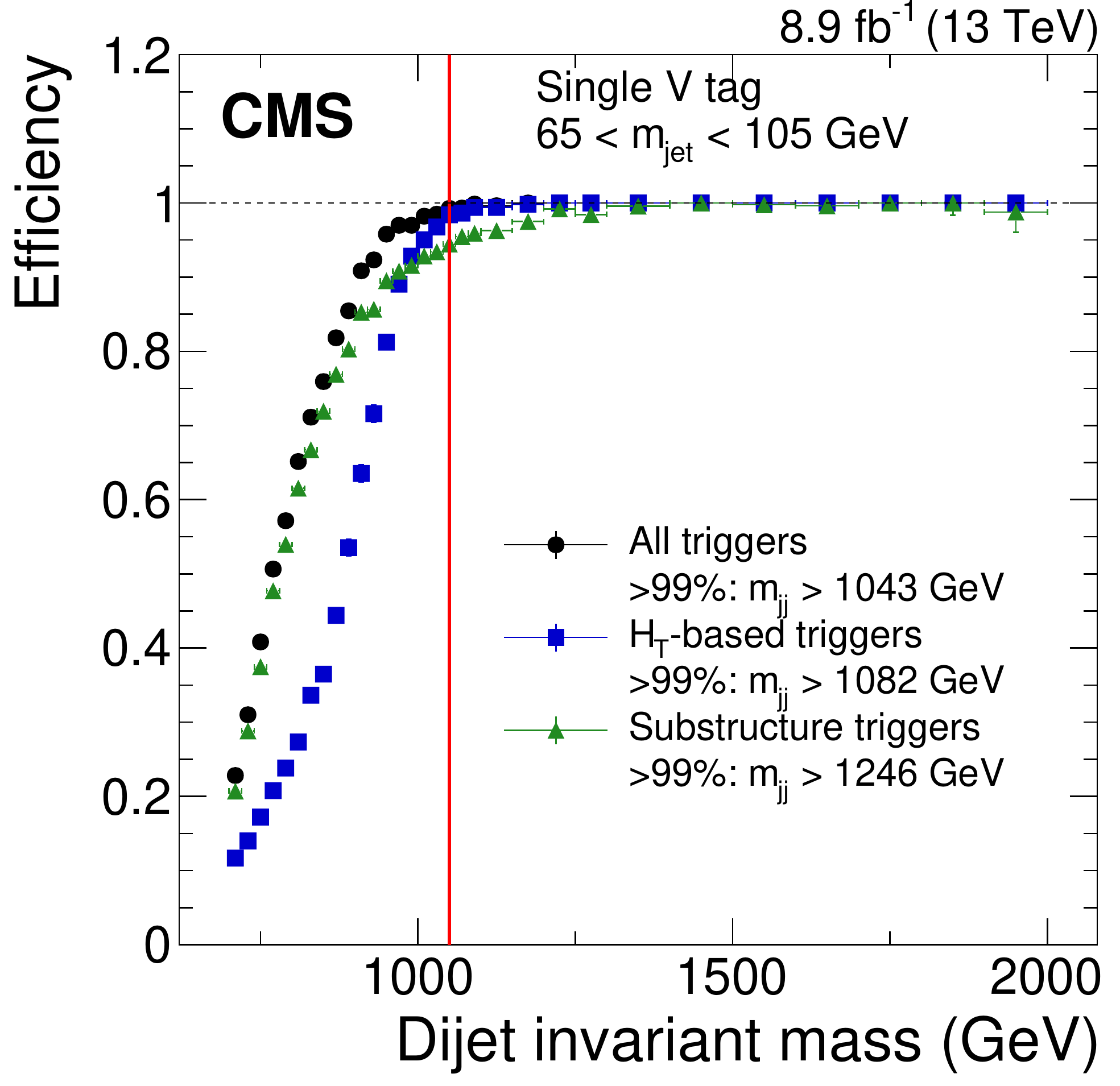}
\hspace{0.1\cmsFigWidth}
\includegraphics[width = \cmsFigWidth]{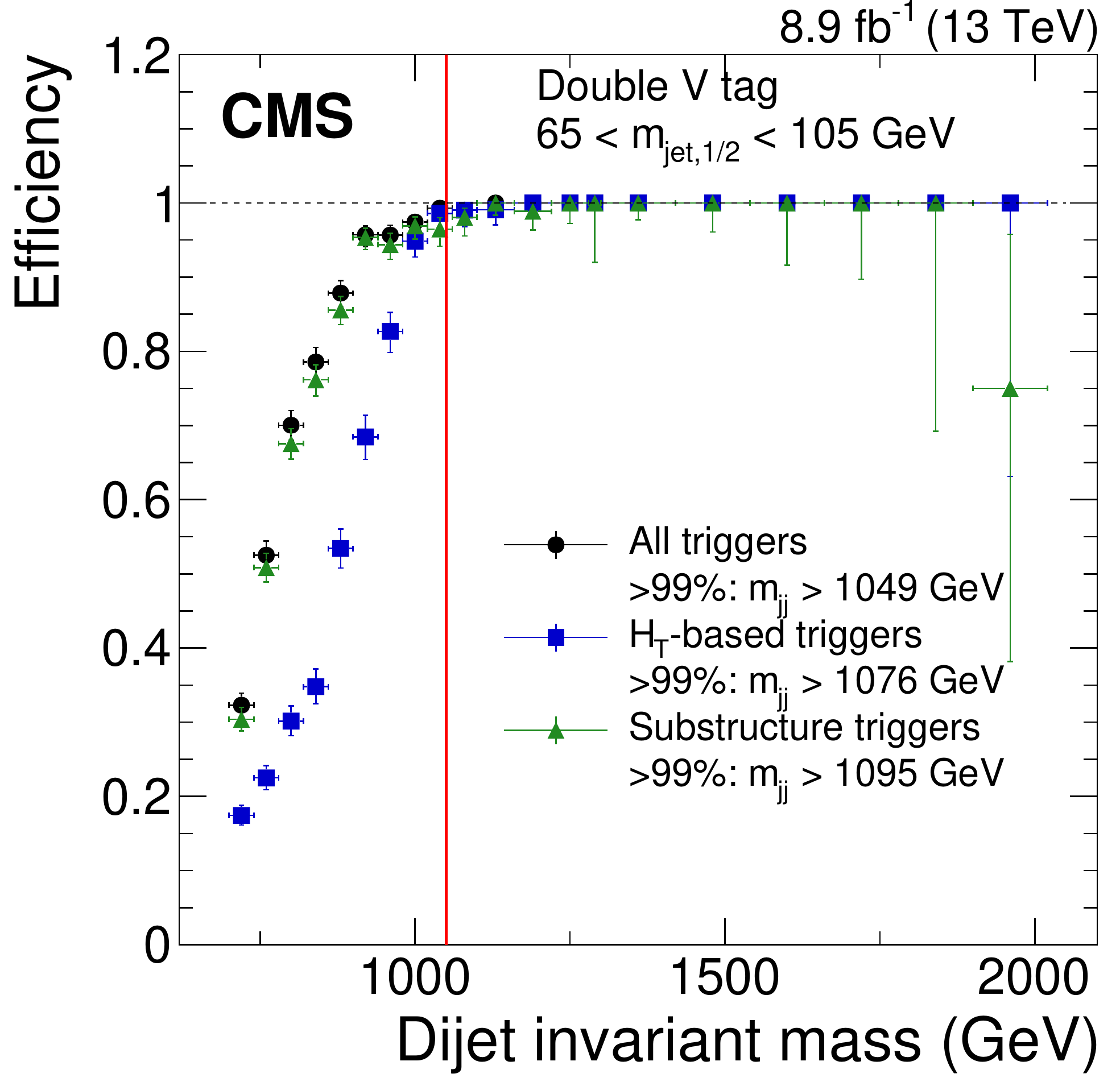}
\caption{Trigger efficiencies for jets passing the inclusive triggers (black), the \HT triggers (blue) or the substructure triggers only (green) as a function of dijet mass for the data-taking period with the highest trigger thresholds. Events are required to contain one jet with a soft-drop mass $m_{jet}$ (left), or two jets with soft-drop masses $m_{jet,1}$ and $m_{jet,2}$ (right), within the signal window of the analysis.
The vertical red line marks the selected threshold value.}
\label{fig:triggerTurnOn}
\end{figure*}

Events are selected online with a range of different jet-based triggers sensitive to
the scalar sum of the transverse momenta of all jets in the event ($\HT$) as well as to the invariant mass of the two leading jets. Additionally, triggers requiring
the presence of one or more jets satisfying loose substructure criteria are used.
Events must satisfy a baseline requirement of $\HT >  800$ or 900\GeV, depending on the instantaneous luminosity during data taking.
Alternatively, a combined requirement of \HT above a threshold of 650--700\GeV and
a single jet with \pt above 360\GeV and trimmed jet mass (as defined in Sec.~\ref{subsec:Vhadr}) \mJ{} $>$ 30--50\GeV qualifies an event to be considered in the analysis.
The combination of triggers is chosen to optimize the value of the dijet invariant mass, \mjj{}, above which the triggers are highly efficient.
The trigger selection reaches an efficiency of at least 99\% for events in which \mjj{} is greater than 1050\GeV and at least one of the two leading-\pt jets has a soft-drop jet mass (as defined in Sec.~\ref{subsec:Vhadr}) above 65\GeV.
The trigger efficiencies comparing substructure and \HT triggers are illustrated in Fig.~\ref{fig:triggerTurnOn} using an orthogonal single muon data set. The individual 99\% efficiency thresholds for events with one and two jets with jet mass above 65\GeV are 1043\GeV and 1049\GeV, respectively. If no jet mass requirements are applied, as is the case for some control distributions in Figure~\ref{fig:DataMCcpl}, the more stringent mass cut of 1080\GeV is used.

Offline, all events are required to have at least one primary vertex reconstructed within a 24\cm window along the beam axis, with
a transverse distance from the nominal $pp$ interaction region of less than
2\cm~\cite{Chatrchyan:2014fea}. The reconstructed vertex with the largest value of summed physics object $\pt^2$ is taken to be the primary $\Pp\Pp$ interaction vertex. The physics objects are the objects returned by a jet finding algorithm~\cite{Cacciari:2008gp,Cacciari:2011ma} applied to all charged tracks associated with the vertex, plus the corresponding associated missing transverse momentum.

\subsection{Substructure variable corrections and validation}\label{subsec:SubVal}

Since discrepancies between data and simulation in the jet substructure variables $\mJ$ and $\nsubj$ could bias the signal efficiency estimated from the simulated samples, the modeling of the signal efficiency is cross-checked in a signal-free sample with jets having characteristics that are similar to those expected for a genuine signal~\cite{Khachatryan:2014vla}.
A sample of high-\pt $\PW$ bosons that decay hadronically and are reconstructed as a single AK8 jet is studied in semileptonic \ttbar and single top quark events. Scale factors for the $\nsubj$ selection efficiency are extracted following the method described in Ref.~\cite{Khachatryan:2014vla}. In this method, a simultaneous fit to the jet mass distributions for different ranges of $\nsubj$ is performed to separate the $\PW$ boson signal from the combinatorial components in the top quark enriched sample, in both data and simulation.
The scale factors are summarized in Table~\ref{tab:WtaggingScaleFactors} and are used to correct the total signal efficiency and the $\VV$ background normalization predicted by the simulation.
The uncertainties quoted on the scale factors for the $\nsubj$ selection include systematic uncertainties due to the simulation of the \ttbar topology (nearby jets, \pt spectrum), computed comparing different combinations of matrix element and shower generators (for details see Ref.~\cite{Khachatryan:2014vla}), and due to the choice of the signal and background fit model.
The $\PW$ jet mass peak position and resolution are also extracted to obtain data versus simulation scale factors for the soft-drop jet mass, as described in Ref.~\cite{JME16003}.
An additional uncertainty to account for the extrapolation to higher momenta of the scale factor obtained from \ttbar samples with jet $\pt \sim 200\GeV$ is calculated, with a resulting factor of $8.5\% \times \ln(\pt/200\GeV)$ for $\nsubj \leq 0.35$ and $65 \leq \mJ \leq 105\GeV$. This uncertainty is estimated based on the difference between \PYTHIA{}8 and \HERWIG{++}~2.7.1~\cite{Bahr:2008pv} showering models. For the $0.35<\nsubj \leq 0.75$ and $65 \leq \mJ \leq 105\GeV$ selection, this uncertainty is $3.9\% \times \ln(\pt/200\GeV)$ and is treated as correlated with the uncertainty for $\nsubj \leq 0.35$.
As the kinematic properties of $\PW$ and $\PZ$ jets are very similar, the same corrections are also used in the case where the $\PV$ jet is assumed to come from a \Zo boson.

\begin{table}[htbp]
\centering
\topcaption{Data versus simulation scale factors for the efficiency of the \nsubj{} selection used in this ana\-lysis, as extracted from a top quark enriched data sample and from simulation.}
\begin{scotch}{cc}
\nsubj{} selection & Efficiency scale factor\\
\hline
$ \x\x\y0 < \nsubj \leq 0.35$              & \SFWTAGHPWPT \\
$0.35 < \nsubj \leq 0.75$       & \SFWTAGLPWPT \\
\end{scotch}
\label{tab:WtaggingScaleFactors}
\end{table}

\subsection{Final event selection and categorization}
\label{sec:finalSelection}

\begin{figure*}[tb]
\centering
\includegraphics[width=\cmsFigWidth]{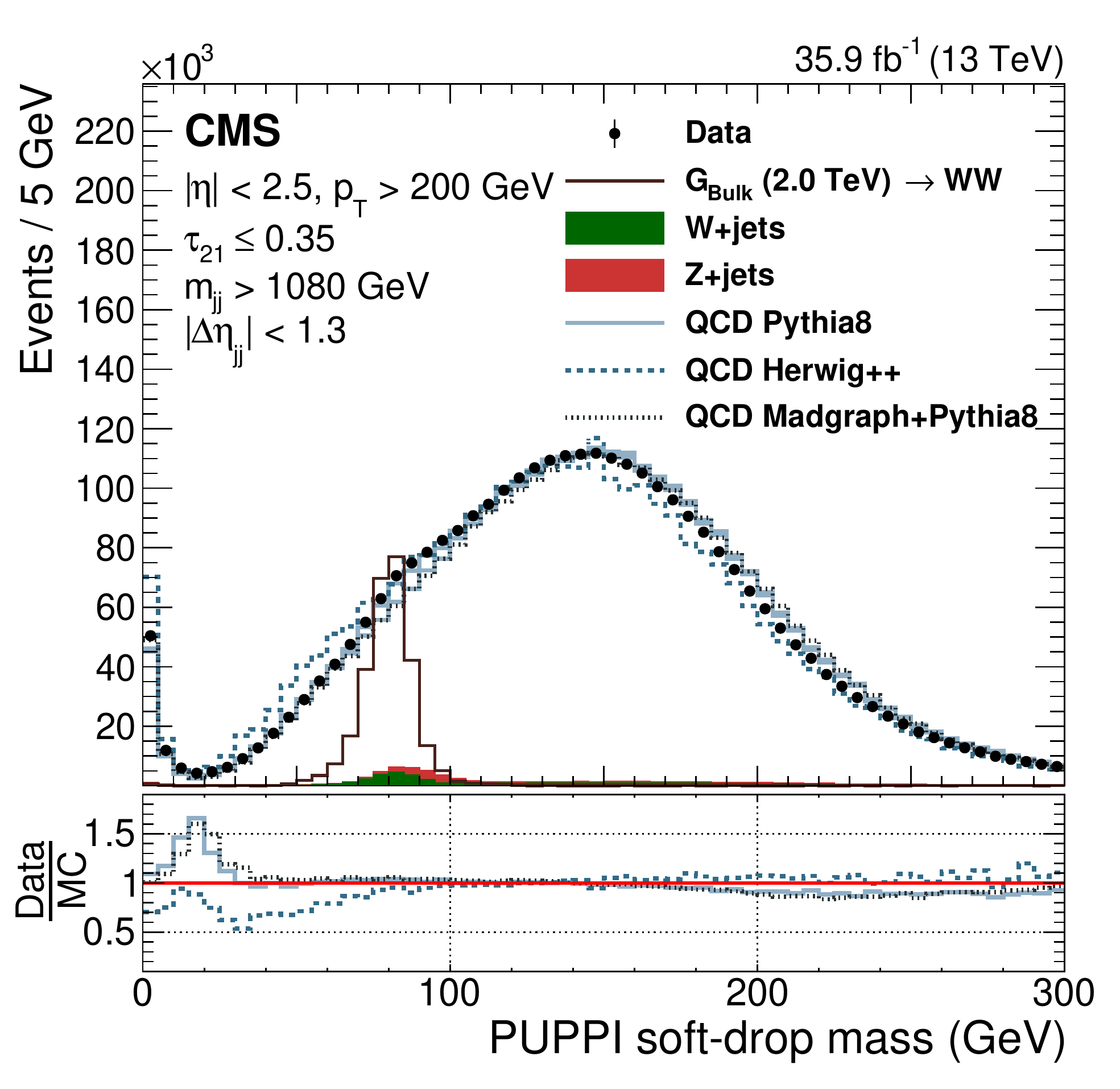}
\hspace{0.1\cmsFigWidth}
\includegraphics[width=\cmsFigWidth]{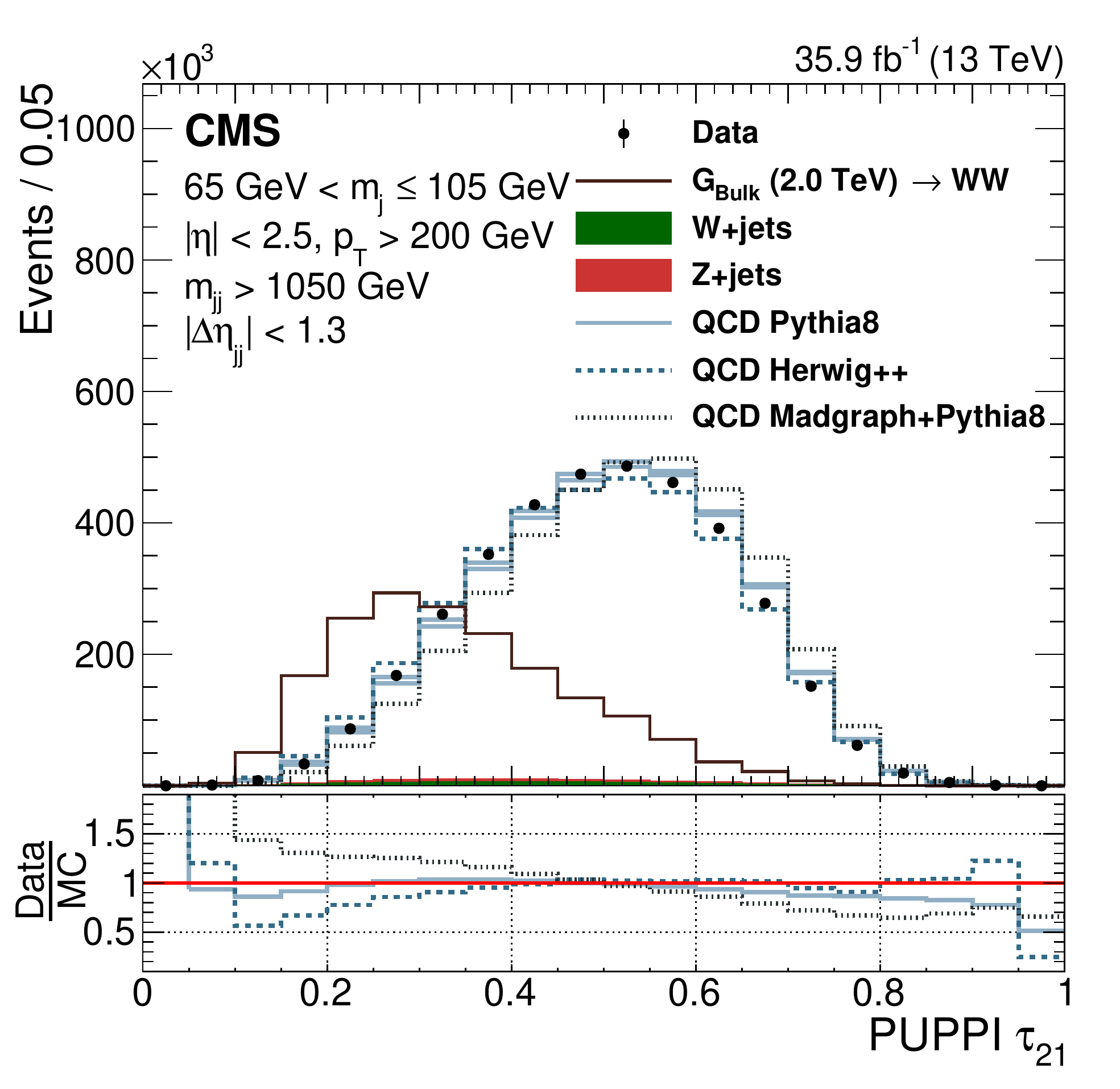}
\caption{
The PUPPI soft-drop jet mass distribution (left) after preselecting and requiring $\nsubj<0.35$, and the PUPPI N-subjettiness $\nsubj$ distribution (right) for data and simulated samples after preselection and requiring a soft-drop mass of $65 \leq \mJ \leq 105\GeV$. The multijet production is shown for three different event generators. The $\PW$+jets and $\PZ$+jets events are stacked with the multijet sample generated with \PYTHIA{}8.
For the PUPPI soft-drop jet mass distribution, the \mjj{} requirement has been raised from the analysis threshold of 1050\GeV to 1080\GeV, since no requirements on the jet mass are applied. The lower subplots show the data over simulation ratio per bin.
}
\label{fig:DataMCcpl}
\end{figure*}

After reconstructing the vector bosons as $\PV$-tagged AK8 jets, we apply the final selections used for the search. For the excited quark search the selections of the $\VV$ case are loosened so that the quark jet candidate is not subjected to a groomed mass or substructure requirement. Any $\PV$ boson candidate, as well as the q jet candidate for the $\PQq\PV$ analysis, must have $\pt> 200$\GeV. If more than two such candidates are present in the event, which is the case for approximately 16\% of selected events, the two jets with the highest $\pt$ are selected. The event is rejected if at least one of the two jets has an angular separation $\Delta R$ smaller than 0.8 from any electron or muon in the event, to allow future use of the results in a combination with studies in the semi- or all-leptonic decay channels \cite{Khachatryan:2014xja,Khachatryan:2016zqb}. Leptons used for this veto need to have a \pt greater than 35 (30)\GeV, an absolute pseudorapidity smaller than 2.5 (2.4), and pass identification criteria that were optimized for high-momentum electrons (muons)~\cite{Khachatryan:2016zqb}.  In addition, we require the two jets to have a separation of $\abs{\Delta\eta_{jj}} < 1.3$ to reject multijet background, which typically contains jets widely separated in $\eta$. Furthermore, \mjj{} must be above 1050\GeV in order to be on the trigger plateau.
Fig.~\ref{fig:DataMCcpl} shows the distribution of the soft-drop jet mass and N-subjettiness variable for the leading jet in the event after this initial selection.

To enhance the analysis sensitivity, the events are categorized according to the characteristics of the $\PV$ jet.
The $\PV$ jet is deemed a $\PW$ boson candidate if its soft-drop mass falls into the range
\SRLOW--$\SRMIDDLE\GeV$,
while it is deemed a $\PZ$ boson candidate if it falls into the range
\SRMIDDLE--$\SRHIGH\GeV$. This leads to three mass categories ($\WW$,  $\WZ$, and  $\ZZ$) for the double-tag analysis and two mass categories ($\PQq\PW$ and $\PQq\PZ$) for the single-tag analysis.
Owing to jet mass resolution effects, up to 30\% of W/Z bosons are reconstructed in the Z/W mass window. For this reason, all three (two) mass categories are considered for all signal categories in the double-tag (single-tag) analysis, respectively.
We select high-purity (HP) $\PV$ jets by requiring $\nsubj \leq 0.35$, and low-purity (LP) $\PV$ jets by requiring $0.35 < \nsubj < 0.75$.
The threshold of 0.35 is chosen to gain significance for mass points below 2.5 (2.2) TeV in the double- (single-) tag region, where the significance achieved with this selection is within 10\% of the maximal significance attained using the optimal selection value for each mass point.
The threshold of 0.75 is chosen to reject less than 1\% of signal events so that the expected significance at high invariant masses is close to maximal.
Events with just one $\PV$ tag are classified according to these two categories. For the double-tag analysis, events are always required to have one HP $\PV$ jet, and are divided into HP and LP events, depending on whether the other
$\PV$ jet is of high or low purity.
Although it is expected that the HP category dominates the total sensitivity of the analysis, the LP category is retained since it provides improved signal efficiency with acceptable background contamination at high resonance masses.
The final categorization in $\PV$ jet purity and $\PV$ jet mass category ($\WW$,  $\WZ$,  $\ZZ$, $\PQq\PW$, and $\PQq\PZ$) yields a total of six orthogonal classes of events for the double-tag analysis and four classes of events for the single-tag analysis.

The two boson (boson and quark jet) candidates, are then combined into a diboson (boson-quark) candidate; the presence of signal events could then be inferred from the observation of localized excesses in the \mjj{} distribution.

\section{Modeling of background and signal}
\label{sec:backgroundestimation}

\subsection{Signal modeling}
\label{sec:signal}

\begin{figure*}[htb]
\centering
\includegraphics[width=\cmsFigWidth]{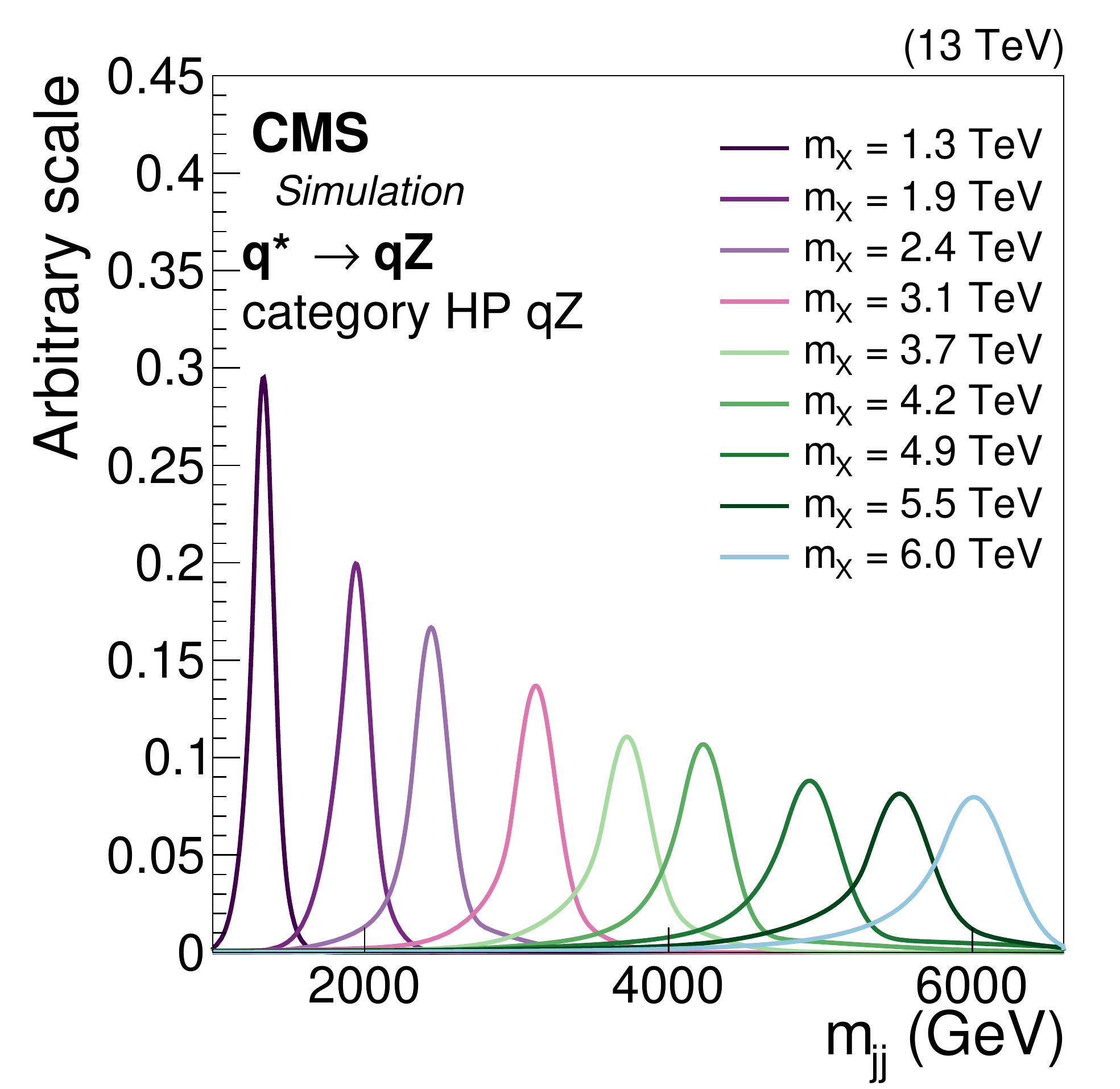}
\hspace{0.1\cmsFigWidth}
\includegraphics[width=\cmsFigWidth]{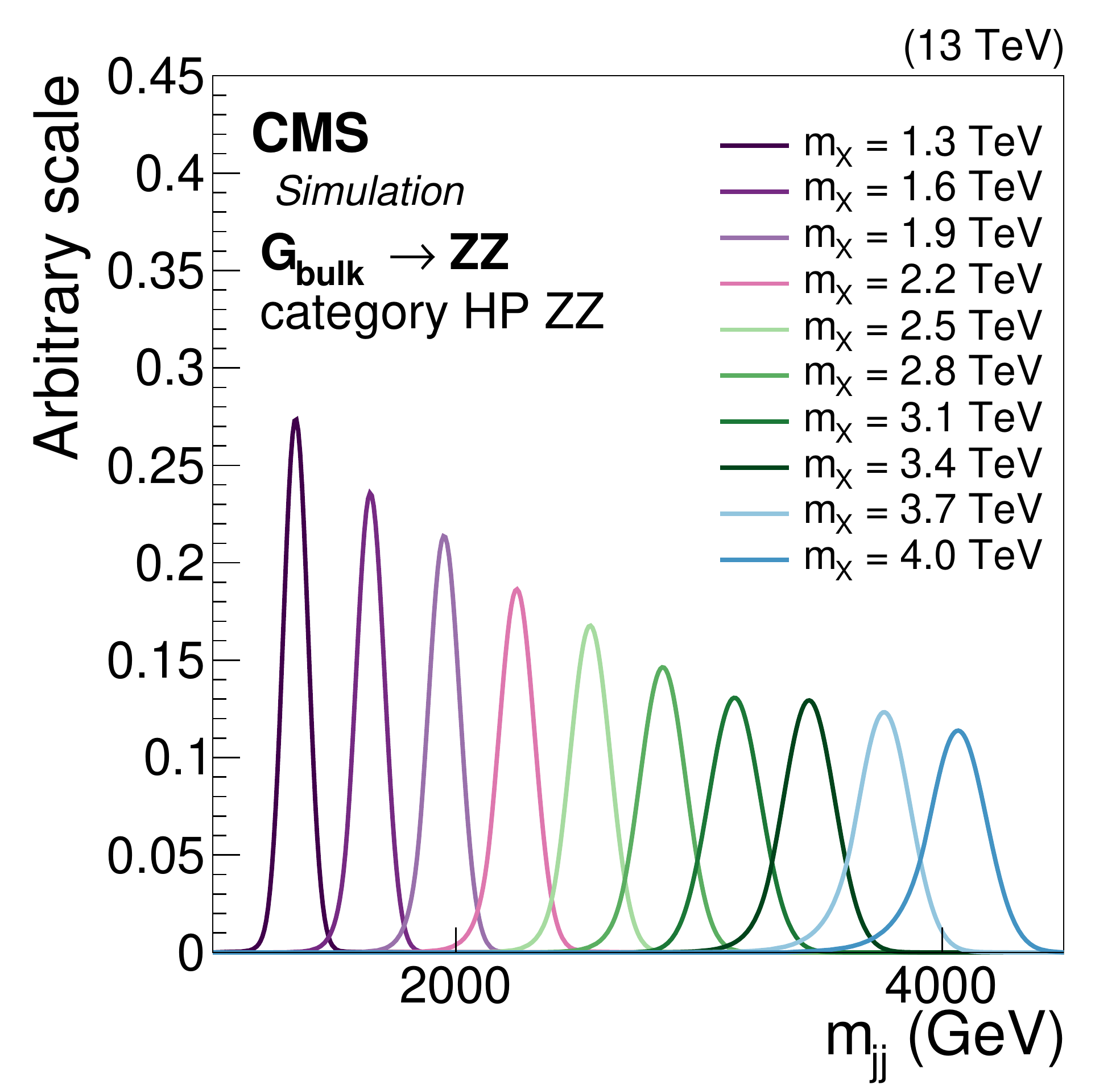}
\caption{Dijet invariant mass distribution for different signal mass hypotheses of the $\PQq^*\to \PQq\Zo$ model (left) and the bulk graviton decaying to a pair of $\PZ$ bosons (right) used to extract the signal shape in the HP category.}
\label{fig:sigfit}
\end{figure*}

Figure~\ref{fig:sigfit} shows the simulated \mjj{} distributions for resonance masses from 1.3 to 6\TeV.
The experimental resolution is about 4\%.
We adopt an analytical description of the signal shape, choosing the sum of a crystal-ball
(CB) function~\cite{CrystalBallRef} (\ie, a Gaussian core with a power law tail to low masses) and a Gaussian function to describe the simulated resonance distributions. The parameters of the analytic shapes are extracted from fits to the signal simulation. Statistical uncertainties in the parameters are negligible.
A cubic spline interpolation between a set of reference distributions (corresponding
to masses of 1.2, 1.4, 1.6, 1.8, 2.0, 2.5, 3.0, 3.5, 4.0, 4.5, 5.0, 5.5, 6.0, and 6.5\TeV) is used to obtain the expected distribution for intermediate values of resonance mass.

\subsection{Multijet background}
\label{sec:dijetmethod}

\par The $\mjj$~distributions observed in data are dominated by SM background processes, which in turn are dominated by multijet production where quark or gluon jets are falsely identified as $\PV$ jets. Additional subdominant backgrounds include $\PW$ and $\PZ$ boson production, top quark pair production, single top quark production, and nonresonant diboson processes. Those backgrounds are estimated from simulation to each contribute less than about 3\% of the total number of background events in the signal region and are therefore not separated in the background estimation.

We assume that the multijet SM background can be described by a smooth, monotonically decreasing distribution, which can be parametrized.
The search is performed by fitting the sum of the analytical functions for background and signal to the whole dijet spectrum in data. Separate fits are made for each signal mass hypothesis and each analysis category, assuming full correlation between the signal normalization parameters and no correlation between background parameters. The shape of the signal function is fixed through a fit of the signal probability distribution function to the interpolated MC simulations, as described in Sec.~\ref{sec:signal}, while the signal normalization is left floating.
Neither data control regions nor simulated background samples are used directly by this method. The background functions are of the form:
\begin{equation}
\label{eq:dijet1}
\ifthenelse{\boolean{cms@external}}
{
\begin{split}
\frac{\rd N}{\rd \mjj}= & \frac{ P_0(1-\mjj/\sqrt{s})^{P_2} } { (\mjj/\sqrt{s})^{P_1} }\:\:\text{(3-par. form)}, \\
\frac{\rd N}{\rd \mjj}= & \frac{ P_0 } { (\mjj/\sqrt{s})^{P_1} }\:\:\text{(2-par. form)},
\end{split}
}
{
\frac{\rd N}{\rd \mjj}= \frac{ P_0(1-\mjj/\sqrt{s})^{P_2} } { (\mjj/\sqrt{s})^{P_1} }\:\:\text{(3-par. form)},
\quad\quad\quad\quad
\frac{\rd N}{\rd \mjj}= \frac{ P_0 } { (\mjj/\sqrt{s})^{P_1} }\:\:\text{(2-par. form)},
}
\end{equation}
where
\mjj{} is the dijet invariant mass (equivalent to the diboson or quark-boson candidate mass \mVV{} or \mqV{} for the signal),
$\sqrt{s}$ is the center-of-mass energy,
$P_0$ is a normalization parameter for the probability density function, and $P_1$ and $P_2$ describe the shape.
Starting from the two-parameter functional form, a Fisher F-test \cite{FisherTest} is used to check at 10\% confidence level,
if additional parameters are needed to model the individual background distribution.
For the $\VV$ categories, the two-parameter functional form is found to describe the data
spectra sufficiently well. The $\PQq\PV$ channels are best described by the three-parameter functional form according to the F-test.
Alternative parameterizations and functions with up to five parameters are also studied as a cross-check.

\begin{figure*}[htbp]
\centering
\includegraphics[width=\cmsFigWidth]{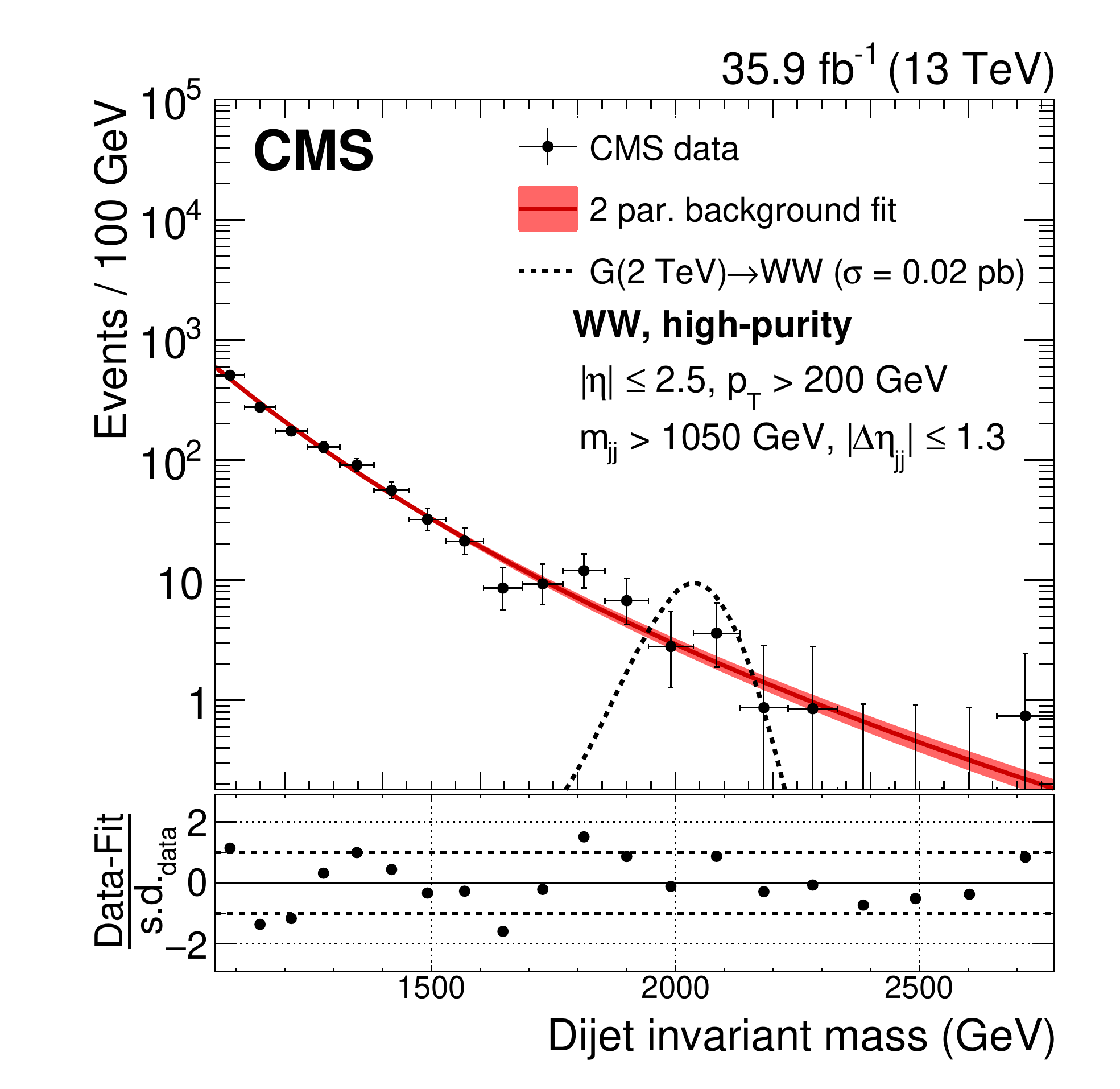}
\hspace{0.1\cmsFigWidth}
\includegraphics[width=\cmsFigWidth]{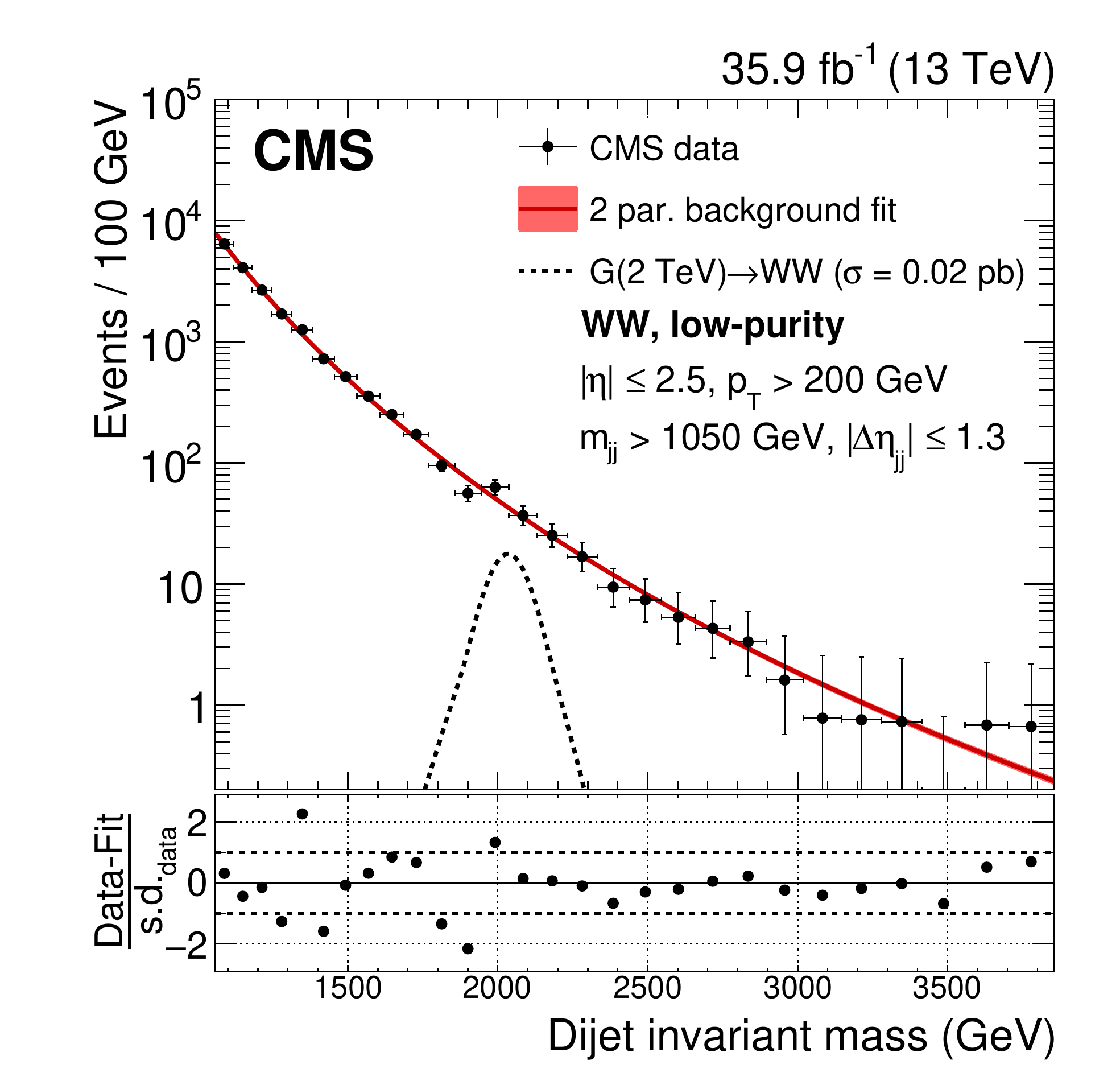}\\
\includegraphics[width=\cmsFigWidth]{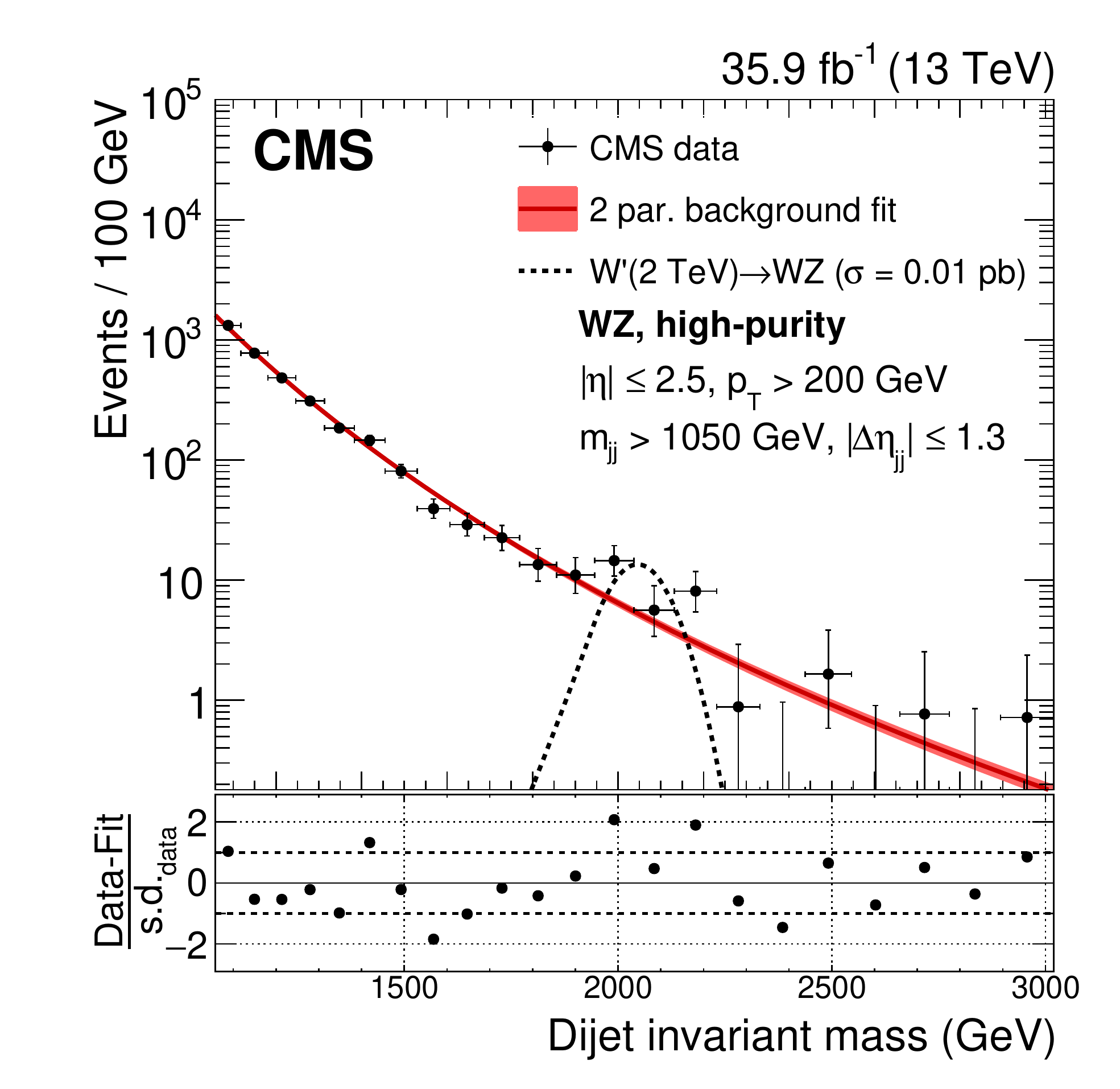}
\hspace{0.1\cmsFigWidth}
\includegraphics[width=\cmsFigWidth]{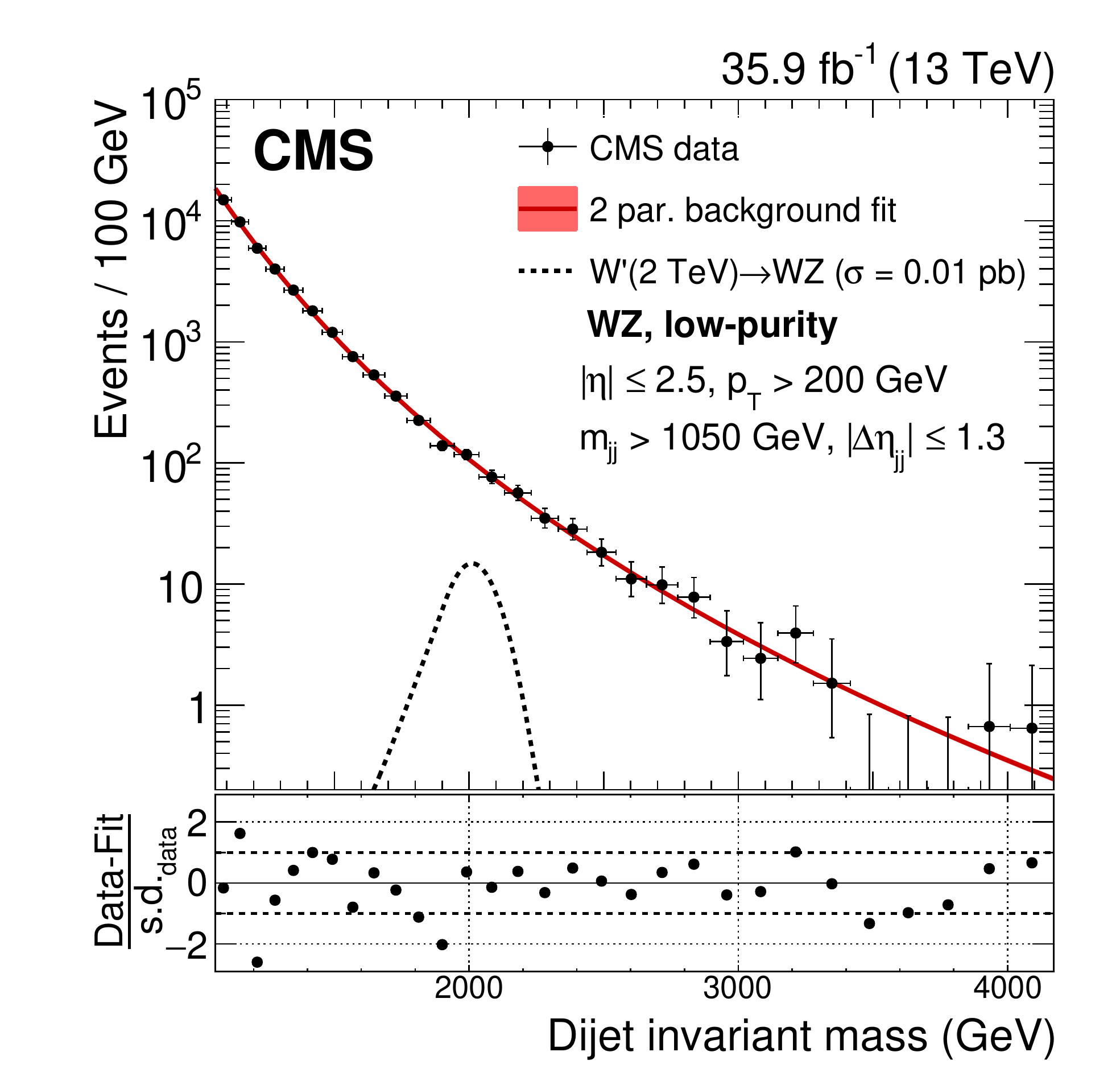} \\
\includegraphics[width=\cmsFigWidth]{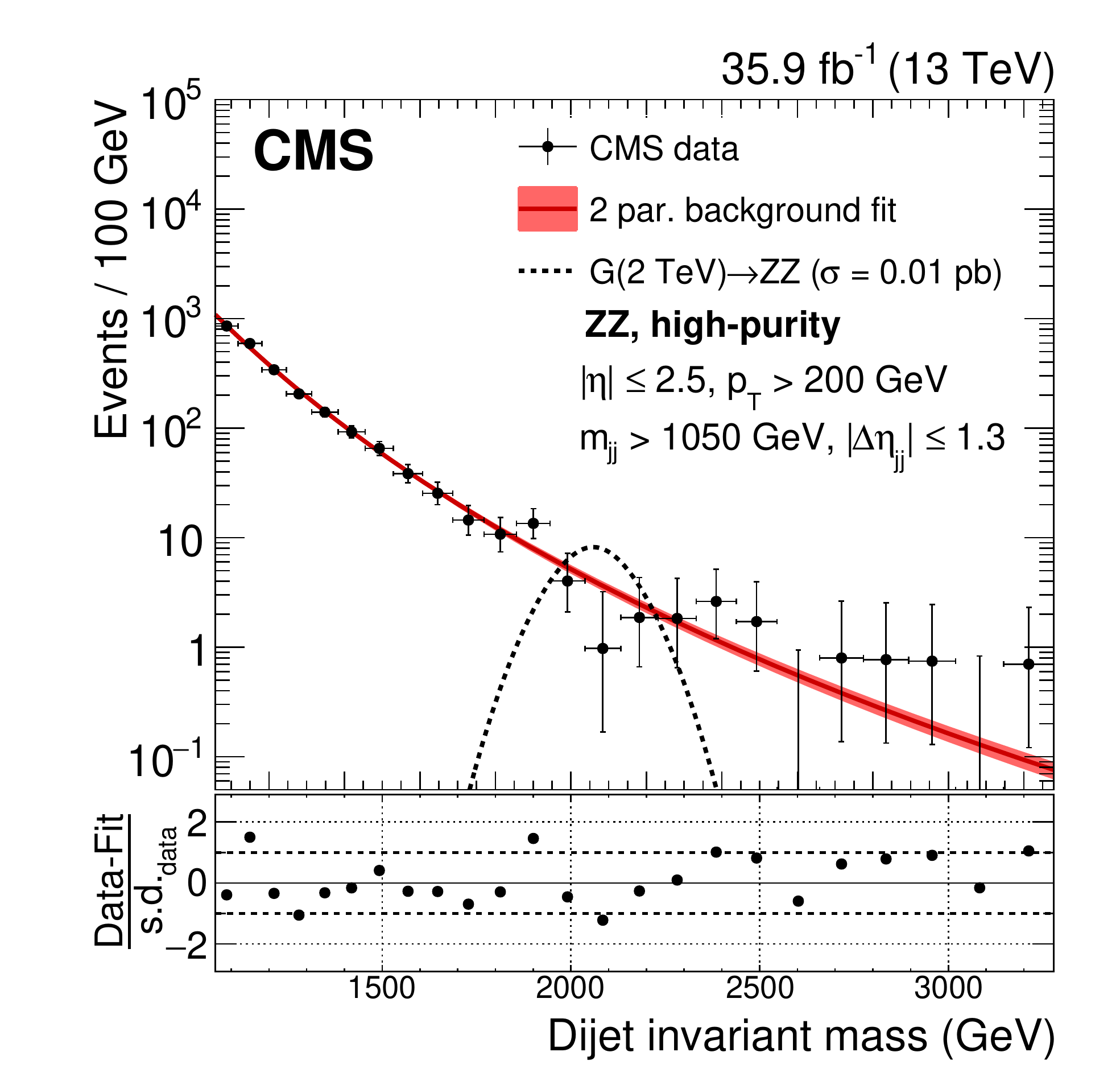}
\hspace{0.1\cmsFigWidth}
\includegraphics[width=\cmsFigWidth]{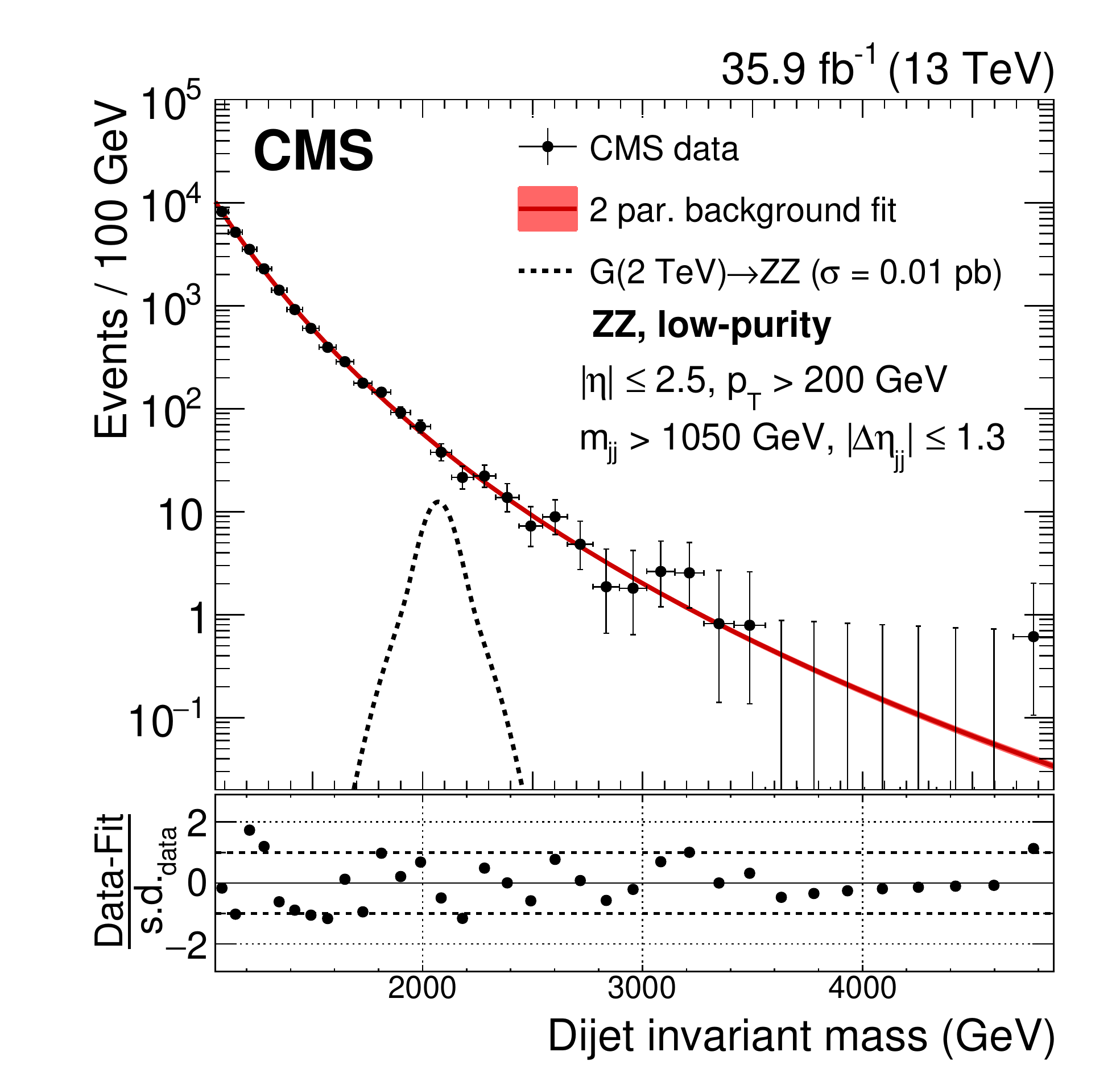}
\caption{The dijet invariant mass distribution \mjj{} in data. On the left, the HP, and on the right, the LP categories are shown for the $\WW$, $\WZ$, and $\ZZ$ categories, from upper to lower.
The solid curve represents a background-only fit to the data distribution where the red shaded area corresponds to the one standard deviation statistical uncertainty of the fit. The dashed line shows the signal shape for a bulk graviton or \PWpr{} of mass 2\TeV.
The lower panels show the corresponding pull distributions, quantifying the agreement between a background-only fit and the data. Note that these fits do not represent the best fit hypotheses used in the statistical analysis where signal-plus-background fits are performed.
}
\label{fig:mjjwithFit_VV}
\end{figure*}

\begin{figure*}[htbp]
\centering
\includegraphics[width=\cmsFigWidth]{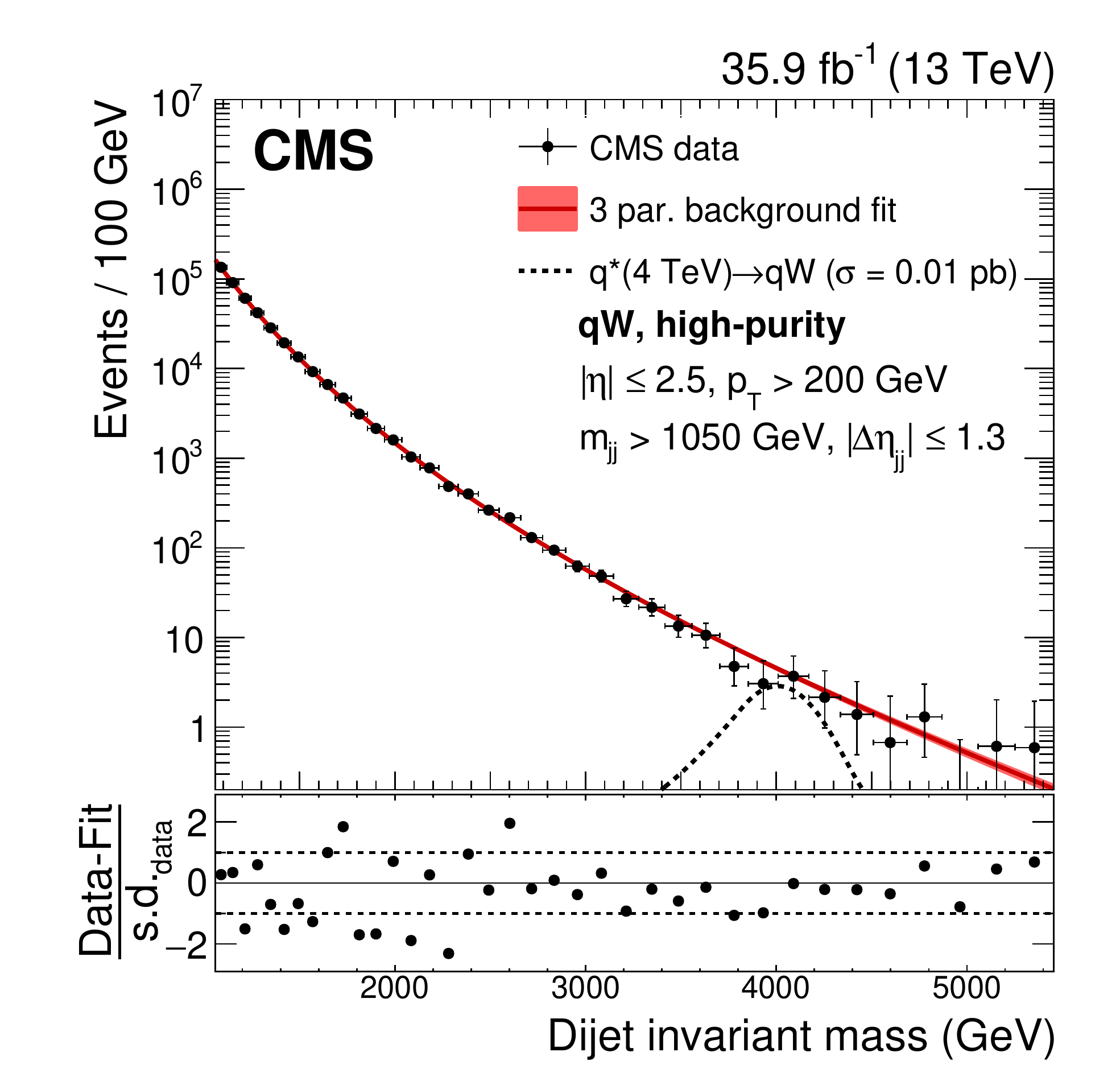}
\hspace{0.1\cmsFigWidth}
\includegraphics[width=\cmsFigWidth]{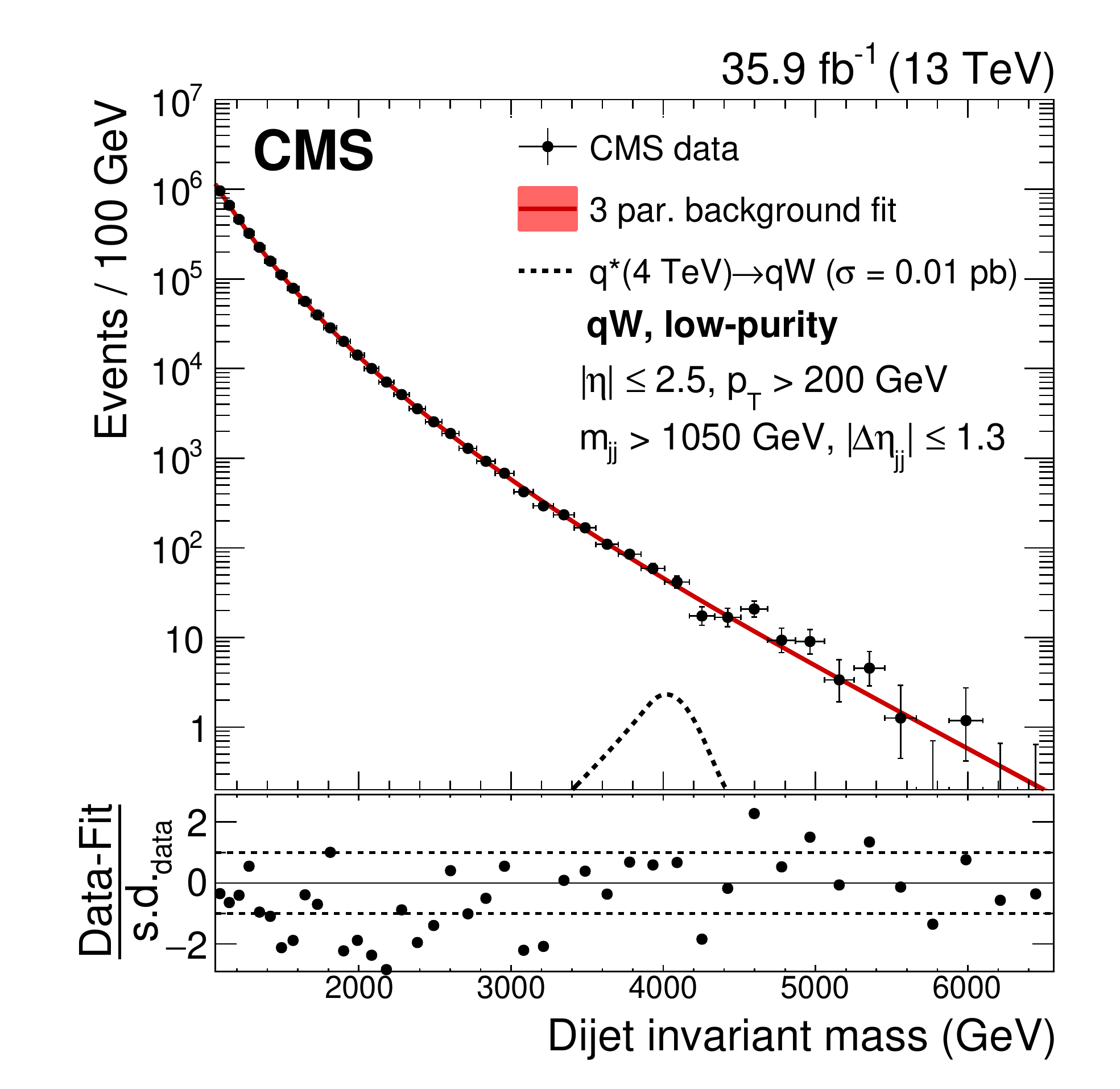}\\
\includegraphics[width=\cmsFigWidth]{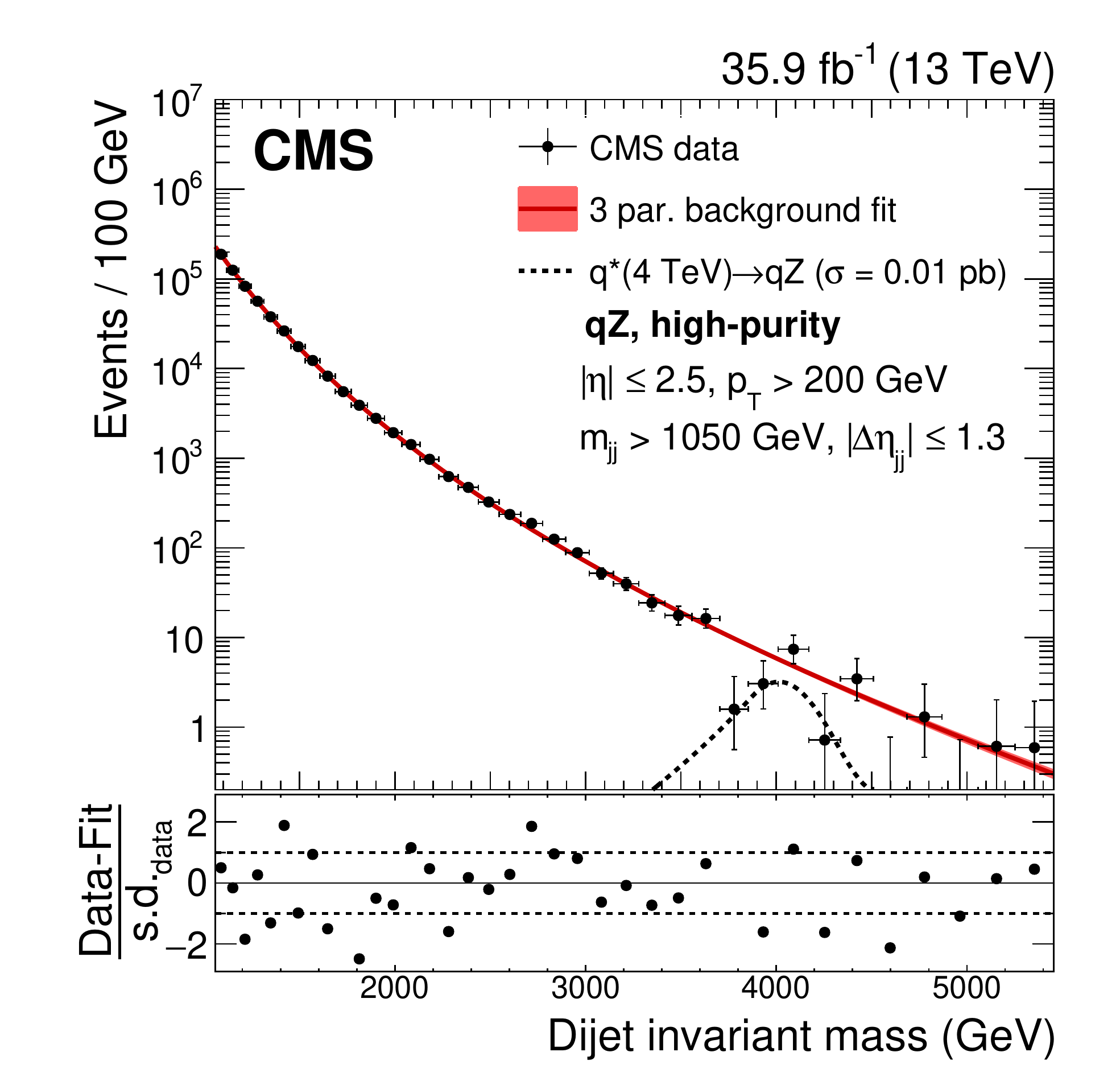}
\hspace{0.1\cmsFigWidth}
\includegraphics[width=\cmsFigWidth]{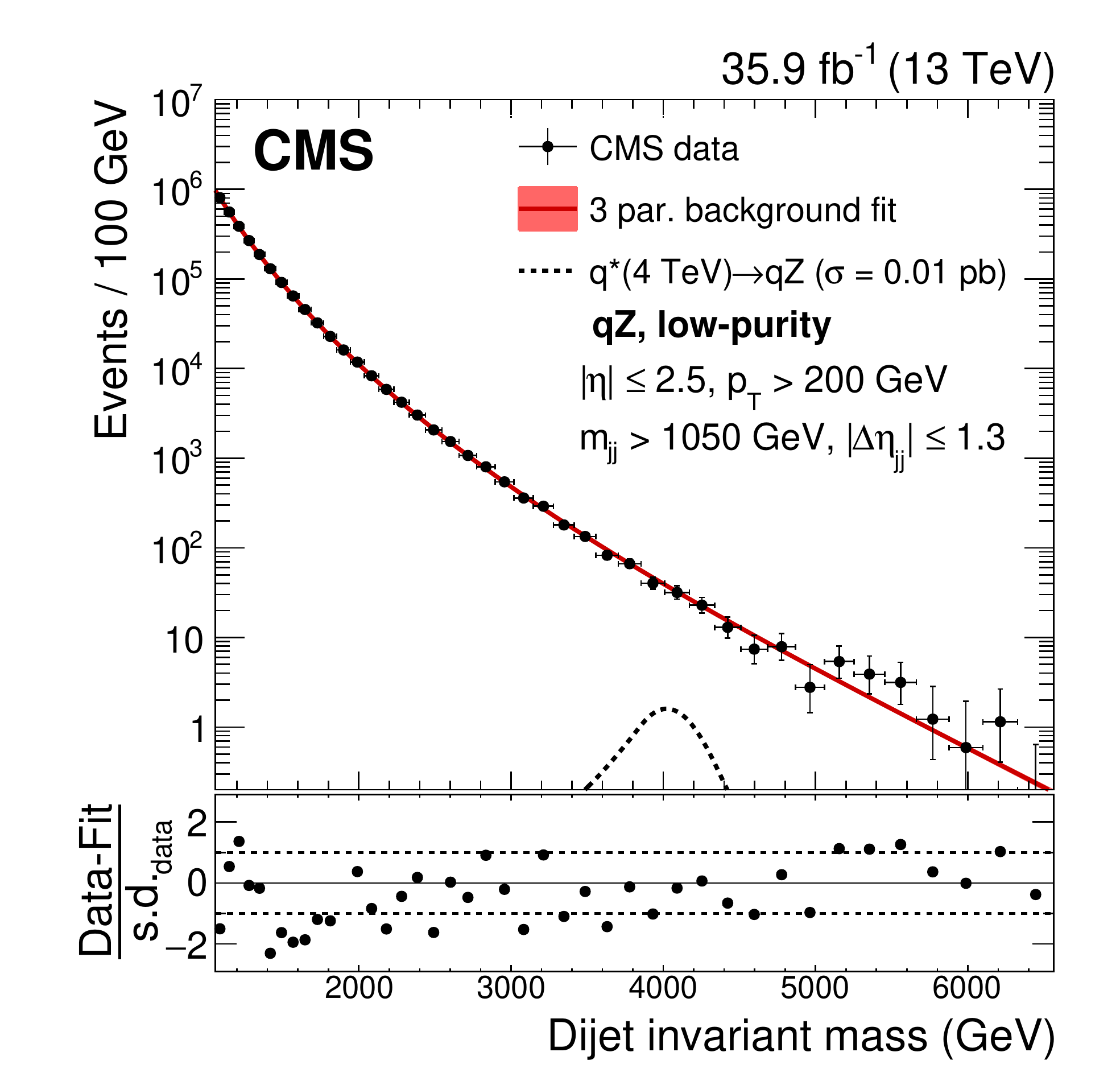}
\caption{The dijet invariant mass distribution \mjj{} in data. On the left, the HP, and on the right, the LP categories are shown for the $\PQq\PW$ and $\PQq\PZ$ categories, from upper to lower.
The solid curve represents a background-only fit to the data distribution where the red shaded area corresponds to the one standard deviation statistical uncertainty of the fit.
The dashed line shows the signal shape for a $\PQq^*$ with a mass of 4\TeV. The lower panels show the corresponding pull distributions, quantifying the agreement between a background-only fit and the data. Note that these fits do not represent the best-fit hypotheses used in the statistical analysis where signal-plus-background fits are performed.
}
\label{fig:mjjwithFit_qV}
\end{figure*}

The fit range is chosen such that it starts where the trigger efficiency has reached its plateau, to avoid any bias from trigger inefficiency, and extends to one bin beyond the bin with the highest \mjj{} event.
The binning~\cite{Khachatryan:2010jd} chosen for the fit follows the detector resolution.
The results of the fits are shown in Figs.~\ref{fig:mjjwithFit_VV} and \ref{fig:mjjwithFit_qV}. The \textit{x-axis} ranges have been chosen to include the most massive observed event in each category, so there are no overflow data.
The solid red curve represents the results of the maximum likelihood fit to the data, with the number of expected signal events fixed to zero.

\section{Systematic uncertainties}
\label{sec:systematicuncertainties}

\subsection{Systematic uncertainties in the background estimation}

\par The background estimation for each signal mass hypothesis is obtained from a fit of the signal plus background function to the full range of
the \mjj{} spectrum. As such, the only relevant uncertainty originates from the covariance matrix of the dijet mass fit function. Different parametrizations of the background fit function were studied and the observed variations of the limit were found to be negligible. This ambiguity in the choice of the background fit function is therefore not considered as an uncertainty source.

\subsection{Systematic uncertainties in the signal prediction}

\par The dominant uncertainty in the signal selection efficiency arises from uncertainties in the $\PV$ tagging efficiency. As described in Sec.~\ref{subsec:SubVal}, the efficiency of the $\PV$ tagging selection is measured in data using a sample enriched in semileptonic \ttbar{} events. A simultaneous fit to that data sample and to a corresponding suitable mixture of simulated top quark-antiquark pair, single top quark and $\PW$+jets events yields both a correction factor to the $\PV$ tagging efficiency in signal samples, as well as a systematic uncertainty in that efficiency, see Table~\ref{tab:WtaggingScaleFactors}.

The signal efficiency and the reconstructed mass shape of the resonance are affected by uncertainties in the jet reconstruction. Jet scale and resolution uncertainties are propagated by rescaling (smearing) the jet properties according to the measured scale (resolution) uncertainties, respectively.
Further, the soft-drop mass is rescaled (smeared) based on the uncertainty in the jet mass scale and resolution.
The selection efficiencies are recalculated on these modified samples, with the resulting changes taken as systematic uncertainties depending on the resonance mass. The induced changes in the reconstructed mass shape of the resonances are propagated as uncertainties in the peak position and width of both the Gaussian core of the CB function and the Gaussian function used in the signal parametrization. Additionally, the induced relative migration among $\PV$ jet mass categories is evaluated, but this does not affect the overall signal efficiency.

\par The uncertainty in the knowledge of the integrated luminosity of the data sample (\LUMIUNCERT)~\cite{CMS-PAS-LUM-17-001} introduces an uncertainty in the number of signal events passing the final selection.

We evaluate the influence of uncertainties in the PDFs and the choice of factorization ($\mu_{\mathrm{F}}$) and renormalization ($\mu_{\mathrm{R}}$) scales on the signal cross section and acceptance by considering
differences in the predicted kinematics of the resonance.
Acceptance and signal cross section effects are treated separately: while the signal acceptance uncertainty
is taken into account in the statistical analysis, the signal cross
section uncertainty is instead considered as an uncertainty in the
theory cross section.
The NNPDF 3.0~\cite{Ball:2014uwa} LO set of PDFs is used to estimate PDF uncertainties.
Following Refs.~\cite{Cacciari:2003fi,Catani:2003zt}, we evaluate the uncertainties in the signal prediction due to missing higher order calculations by varying the default choice of scales in the following six combinations of factors: $(\mu_{\mathrm{F}}$, $\mu_{\mathrm{R}})$ $\times$ $(1/2, 1/2)$, $(1/2, 1)$, $(1,1/2)$, $(2, 2)$, $(2, 1)$, and $(1, 2)$.
The resulting cross section uncertainties vary from 4 to 72\% and from 2 to 23\%, respectively, depending on the resonance mass, particle type, and its production mechanism.
The uncertainty in the signal acceptance from the choice of PDFs and of factorization and renormalization scales ranges from 0.1 to 2\% and $<$0.1\%, respectively.
In addition, the impact of PDF variations on the signal shape are evaluated and propagated as uncertainties in the signal width and peak position, analogously to the treatment of shape uncertainties for jet energy-momentum scale and resolution.
Table~\ref{tab:VV_systematicssummary_signal} summarizes the systematic uncertainties considered in the statistical analysis.

\begin{table*}[!tb]
\topcaption{Summary of the signal systematic uncertainties for the analysis and their impact on the event yield in the signal region and on the
reconstructed \mjj~shape (mean and width). The jet mass and $\PV$ tagging uncertainties result in migrations between event categories. The effects of the PDF and scale uncertainties in the signal cross section are not included as nuisance parameters in the limit setting procedure, but are assigned to the theory predictions.}
\centering
\ifthenelse{\boolean{cms@external}}{}{\resizebox{\textwidth}{!}}
{
\begin{scotch}{lccccc}
\multirow{3}{*}{Source}                         & \multirow{3}{*}{Relevant quantity}      & \multicolumn{4}{c}{\x\x Uncertainty (\%)}\\
                        &      & \multicolumn{2}{c}{Double-tag} & \multicolumn{2}{c}{Single-tag} \\
&                        & HP+HP  & HP+LP & HP+j & LP+j \\
\hline
Jet energy scale                 & Resonance shape        & 2      & 2   & 2  & 2  \\
Jet energy resolution            & Resonance shape        & 6      & 7   & 4  & 3  \\
PDF                              & Resonance shape        & 5      & 7   & 13  & 8  \\
\hline
Jet energy scale                 & Signal yield           & \multicolumn{2}{c}{$<$1}& \multicolumn{2}{c}{$<$1}\\
Jet energy resolution            & Signal yield           & \multicolumn{2}{c}{$<$1}& \multicolumn{2}{c}{$<$1}\\
Jet mass scale                   & Signal yield           & \multicolumn{2}{c}{$<$2}& \multicolumn{2}{c}{$<$1}\\
Jet mass resolution              & Signal yield           & \multicolumn{2}{c}{$<$6}& \multicolumn{2}{c}{$<$8}\\
Pileup                           & Signal yield           & \multicolumn{4}{c}{\x\x 2}\\
PDF (acceptance)                 & Signal yield           & \multicolumn{4}{c}{\x\x 2}\\
Integrated luminosity            & Signal yield           & \multicolumn{4}{c}{\x\x 2.5}\\
\hline
Jet mass scale                   & Migration              & \multicolumn{2}{c}{\x$<$36}& \multicolumn{2}{c}{\x$<$10}\\
Jet mass resolution              & Migration              & \multicolumn{2}{c}{\x$<$25}& \multicolumn{2}{c}{$<$7}\\
V tagging \nsubj{}               & Migration              & 22 & 33 & 11   & 22 \\
V tagging \pt-dependence         & Migration              & 19--40 & 14--29 & \x9--23   & \x4--11 \\
\hline
PDF and scales (\PWpr and \PZpr) & Theory                 & \multicolumn{2}{c}{\x2--18}& & \\
PDF and scales (\BulkG)          & Theory                 & \multicolumn{2}{c}{\x8--78}& & \\
PDF and scales (q*)              & Theory                 & &       & \multicolumn{2}{c}{\x1--61}\\
\end{scotch}
}
\label{tab:VV_systematicssummary_signal}
\end{table*}

\section{Statistical interpretation}
\label{sec:statisticalinterpretation}

The compatibility between the \mjj~distribution observed in data and the smoothly falling function modeling the standard model background is used to test for the presence of
narrow resonances decaying to two vector bosons or to a vector boson and a quark. We follow
the modified frequentist prescription (asymptotic $\mathrm{CL}_\mathrm{S}$ method) described in Refs.~\cite{CLs1,Junk:1999kv,AsymptCLs}.
The limits are computed using a shape analysis of the dijet invariant mass spectrum. Systematic uncertainties
are treated as nuisance parameters and profiled in the statistical interpretation using log-normal priors, while Gaussian priors are used for shape uncertainties.

\subsection{Limits on narrow-width resonance models}
\label{subsec:narrow-results}

Exclusion limits are set for resonances that arise in the bulk graviton model and in the HVT model~B and for excited quark resonances,
under the assumption of a natural width negligible with respect to the experimental resolution (narrow-width approximation).

\begin{figure*}[tb]
\centering
\includegraphics[width=\cmsFigWidth]{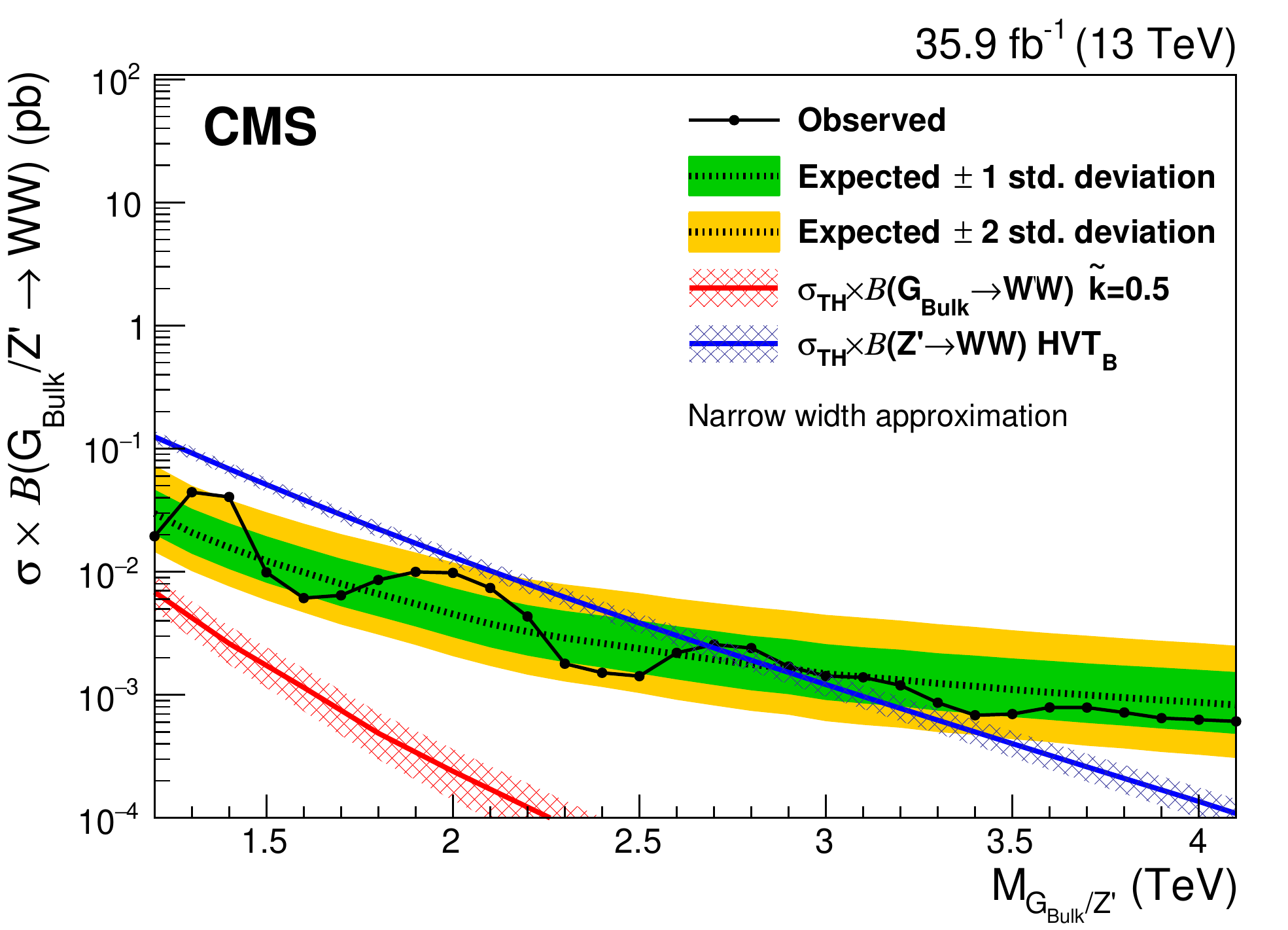}
\includegraphics[width=\cmsFigWidth]{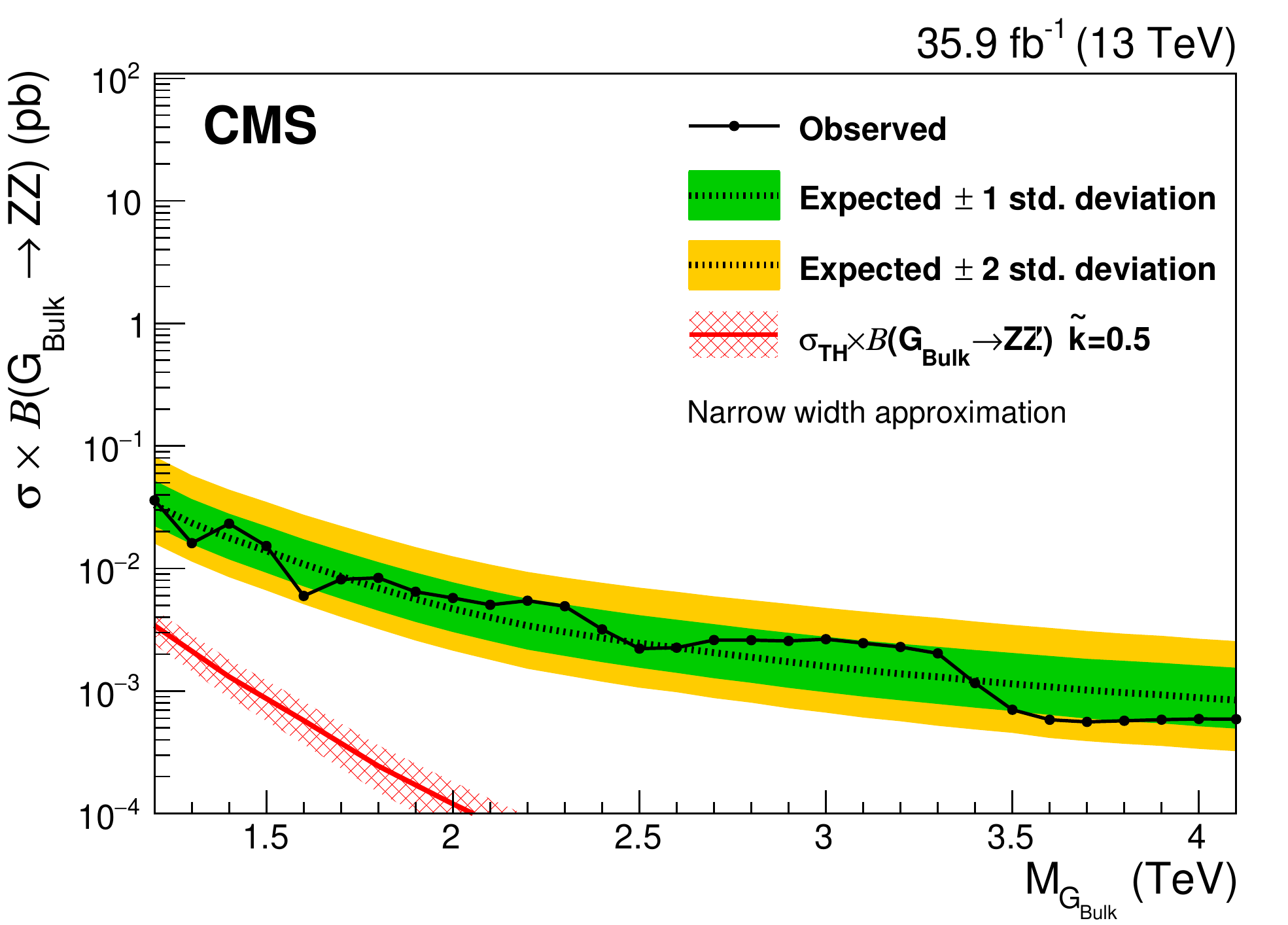}
\hspace{0.1\cmsFigWidth}
\includegraphics[width=\cmsFigWidth]{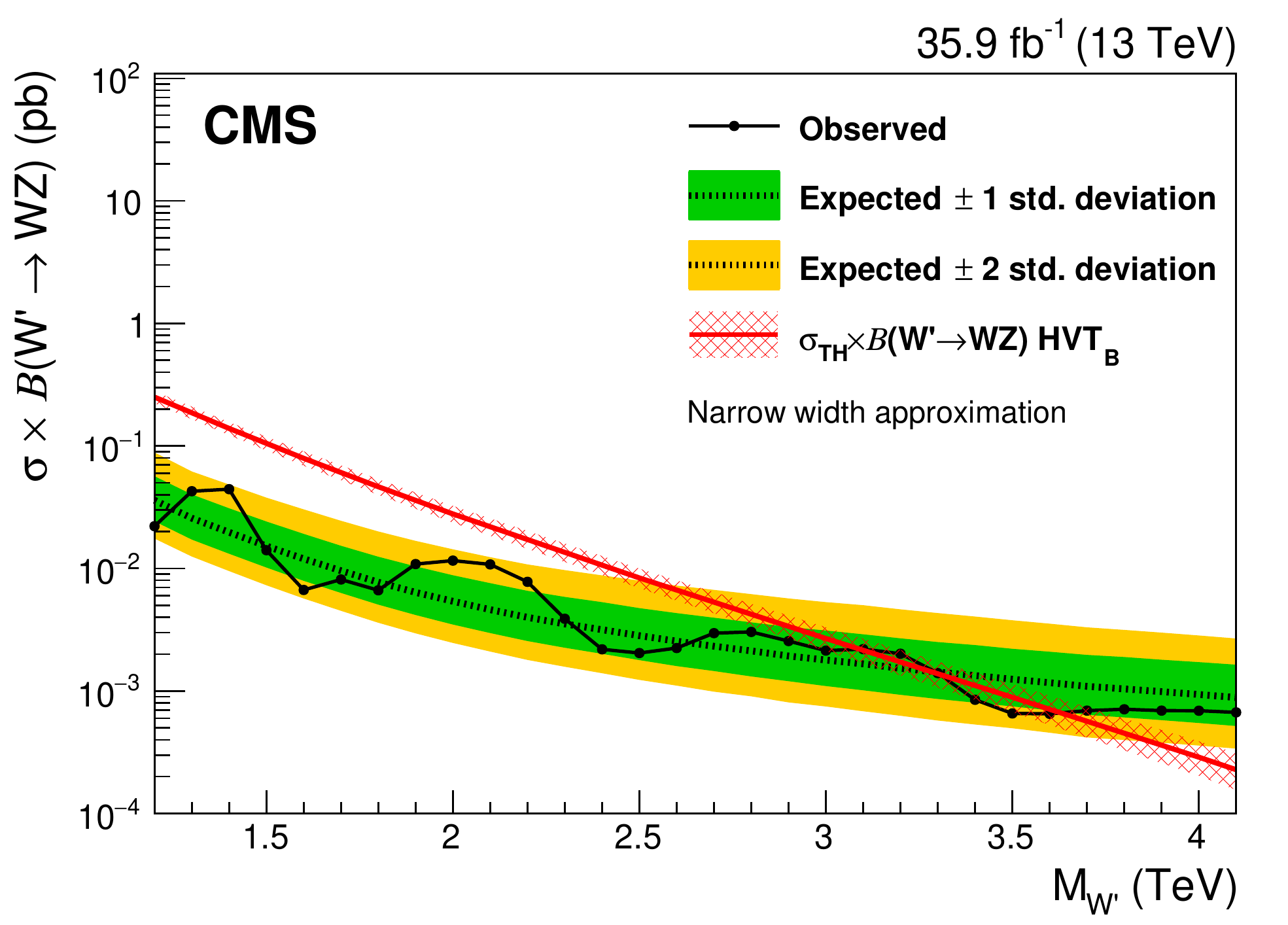}
\includegraphics[width=\cmsFigWidth]{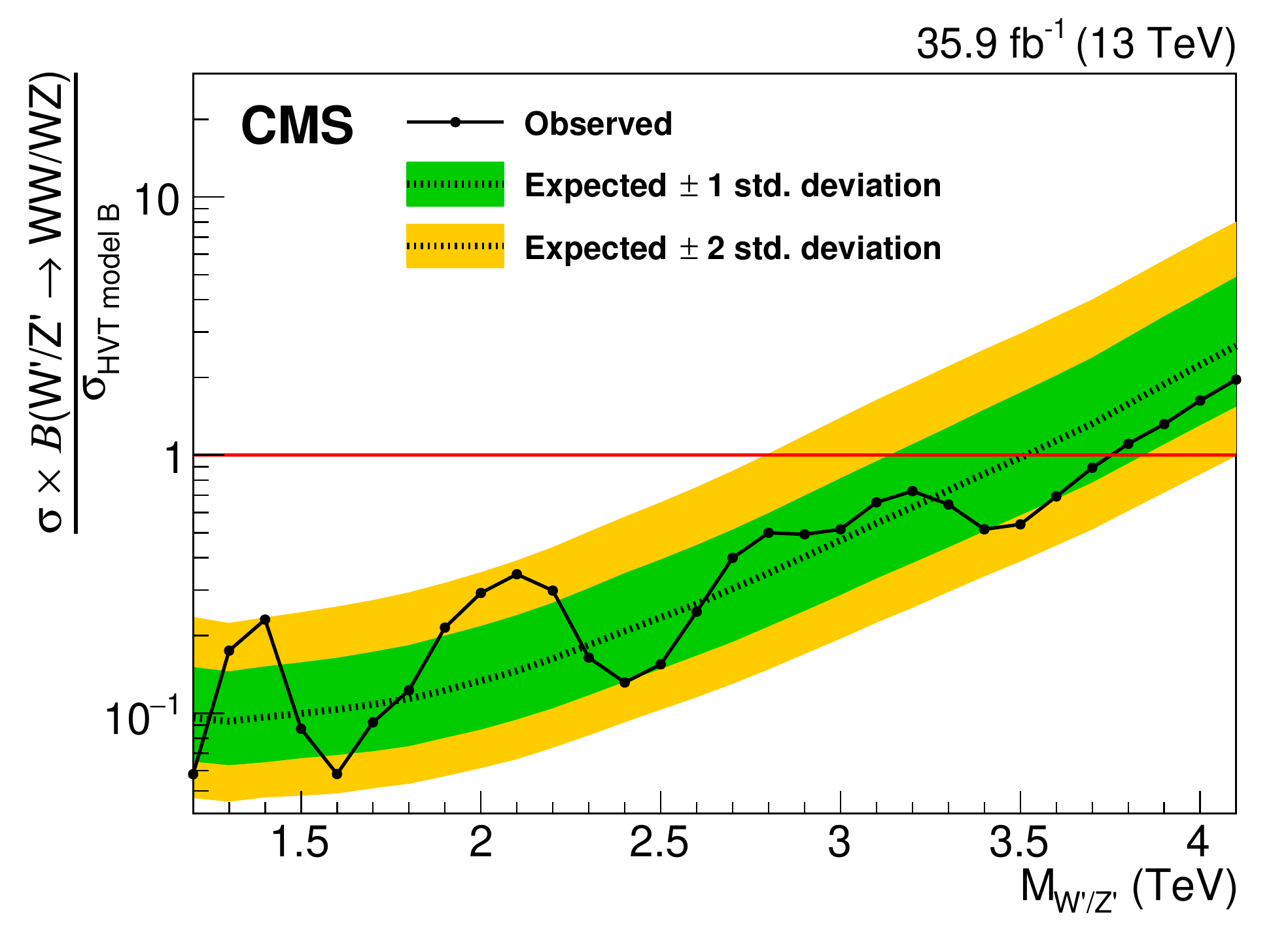}
\caption{Observed (black solid) and expected (black dashed) 95\% C.L. upper limits on the production cross section of a narrow-width resonance decaying to a pair of vector bosons for different signal hypotheses.
Limits are set (upper left plot) on a spin-1 neutral \PZpr and a spin-2 resonance decaying into $\Wo \Wo$, and compared with the prediction of the HVT model~B (blue line) and a bulk graviton model with $\ktilde = 0.5$ (red line).
Limits are also set in the context of a bulk graviton decaying into $\Zo\Zo$ (upper right) with $\ktilde =0.5$ and a spin-1 charged resonance decaying into $\Wo\Zo$ (lower left) and compared with the predictions of the models.
Signal cross section uncertainties are displayed as cross-hatched bands. The plot on the lower right shows the 95\% exclusion bounds on the signal strength for the triplet hypothesis of the HVT model B.}
\label{fig:limitCombined}
\end{figure*}

Figure~\ref{fig:limitCombined} shows the resulting 95\% confidence level (C.L.) expected and observed
exclusion limits on the signal cross section as a function of the resonance mass, for the diboson signal hypotheses.
For a narrow-width spin-2 resonance
the observed exclusion limits on the production cross section range from a cross section limit of 36.0\unit{fb} at a resonance mass of 1.3\TeV to the most stringent cross section limit of 0.6\unit{fb} at resonance masses higher than 3.6\TeV.
In the case of charged (uncharged) spin-1 resonances
the observed exclusion limits range from 44.4 (41.6)\unit{fb} at a mass of 1.4 (1.3)\TeV to 0.7 (0.6)\unit{fb} at high resonance masses.

The limits are compared with the product of the theoretical cross section and the branching fraction to $\Wo\Wo$ or $\Zo\Zo$,
for a bulk graviton with $\ktilde = 0.5$. A comparison is also made with the product of the theoretical cross section
and the branching fraction to $\Wo\Zo$ and $\Wo\Wo$ for spin-1 particles predicted by the HVT model~B for both the singlet (\PWpr or \PZpr) and triplet (\PWpr and \PZpr) hypotheses.
The cross section limits for $\PZpr\to\Wo\Wo$ and $\BulkG\to\Wo\Wo$ are not identical because of the difference in acceptances for the two signals. However, since the acceptance of the \PZpr resonance and the bulk graviton decaying to $\Wo \Wo$ only differ by less than 11\%, the difference of the exclusion limits between the two models is negligible.

For the HVT model B singlet hypothesis we exclude \PWpr resonances below 3.2\TeV and between 3.3 and 3.6\TeV as well as \PZpr resonances below 2.7\TeV.
The signal cross section uncertainties are displayed as a red (blue) checked band and result in an additional uncertainty in the resonance mass limits of 0.15 (0.08)\TeV.
For the triplet hypothesis of the HVT model B, resonances with masses below 3.8\TeV (3.5\TeV expected) are excluded.

\begin{figure*}[htbp]
\centering
\includegraphics[width=\cmsFigWidth]{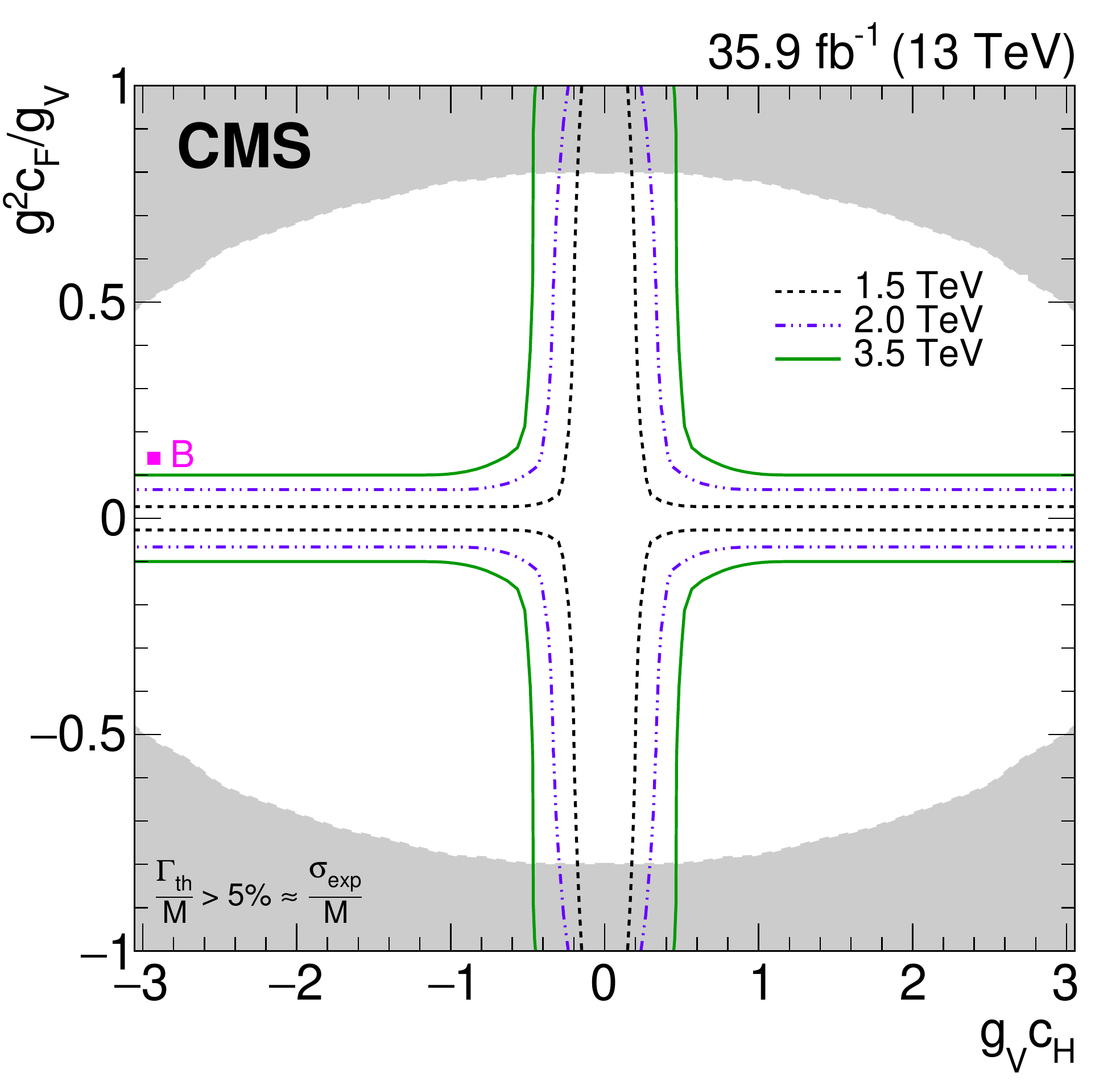}
\caption{Exclusion regions in the plane of the HVT model couplings for three
resonance masses of 1.5, 2, and 3.5\TeV. The point B indicates the values of the coupling parameters used in the benchmark model. The regions of the plane excluded by this search lie outside of the boundaries indicated by the solid and dashed lines. The areas indicated by the solid shading correspond to regions where the theoretical width is larger than the experimental resolution of the present search and the narrow-resonance assumption is therefore not satisfied.}
\label{fig:HVTcouplings}
\end{figure*}

Figure \ref{fig:HVTcouplings} shows a scan of the 95\% C.L. contours in the coupling parameter plane for the triplet hypothesis of the HVT model. The couplings are parametrized in terms of $g_\text{V}c_\text{H}$ and $g^2/g_\text{V}c_\text{F}$, which are related to the coupling of the new resonance to the Higgs boson and to fermions, respectively, as described in Sec.~\ref{sec:simulatedsamples}. Here, $g$ represents the electroweak coupling parameter $g=e/\sin\theta_{\PW}$. The shaded areas indicate the region in the coupling space where the narrow-width assumption is not satisfied.

\begin{figure*}[htbp]
\centering
\includegraphics[width=\cmsFigWidth]{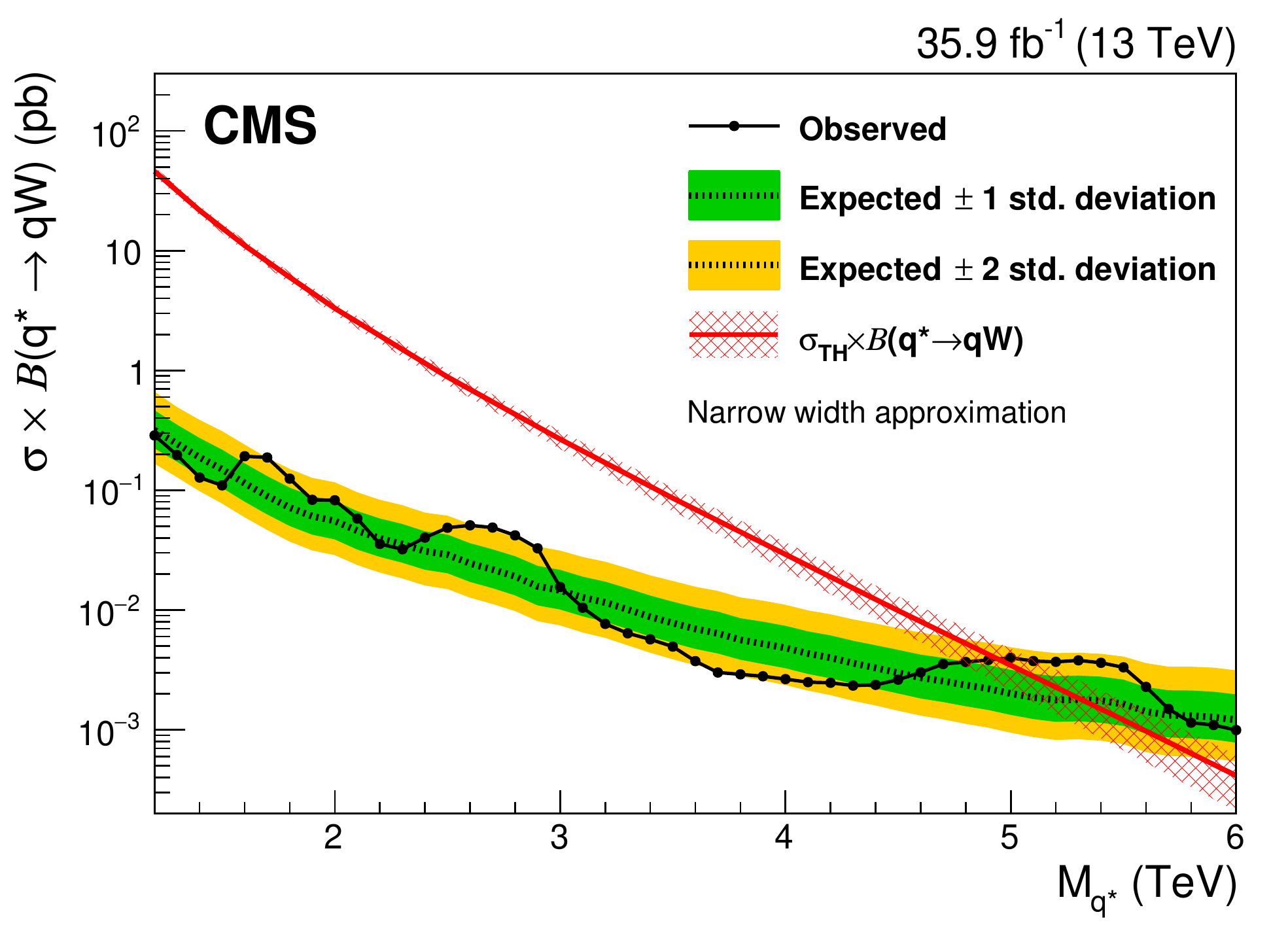}
\hspace{0.1\cmsFigWidth}
\includegraphics[width=\cmsFigWidth]{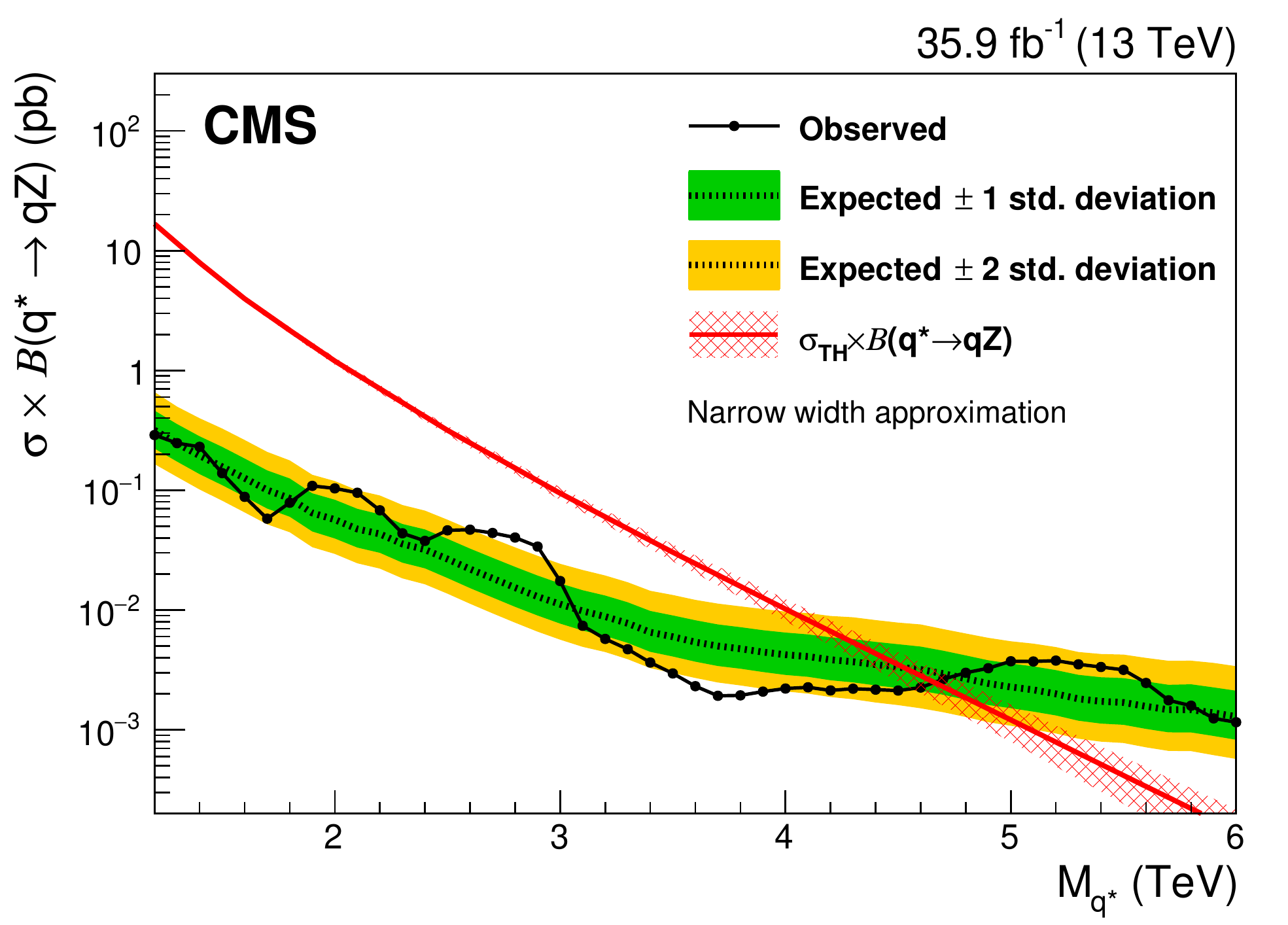}\\
\caption{Observed (black solid) and expected (black dashed) 95\% C.L. upper limits on the production of an excited quark resonance
decaying into $\PQq\PW$ (left) or $\PQq\PZ$ (right) as a function of resonance mass.
Signal cross section uncertainties are displayed as red cross-hatched bands.
}
\label{fig:limitCombined_qV}
\end{figure*}

Figure~\ref{fig:limitCombined_qV} shows the corresponding exclusion limits for excited quarks decaying into $\qo\Wo$ and $\qo\Zo$. The expected cross section limits range from 317\unit{fb} for masses of 1.2\TeV to 1.2\unit{fb} (1.3\unit{fb}) at high resonance masses, while the observed limits cover a range from 287\unit{fb} (289\unit{fb}) to 1.0\unit{fb} (1.2\unit{fb}) between the resonance masses of 1.2 and 6.0\TeV for resonances decaying to $\PQq\PW$ ($\PQq\PZ$).
We exclude excited quark resonances decaying into $\qo\Wo$ and $\qo\Zo$ with masses below 5.0 and 4.7\TeV, respectively.
The signal cross section uncertainties are displayed as a red checked band and result in an additional uncertainty in the resonance mass limits of 0.13--0.20\TeV.

\section{Summary}
\label{sec:summary}

A search is presented for new massive narrow resonances decaying to $\WW$, $\ZZ$, $\WZ$, $\PQq\PW$, or $\PQq\PZ$, in which the bosons decay hadronically into dijet final states.
Hadronic $\PW$ and $\PZ$ boson decays are identified by requiring a jet with mass compatible with the $\PW$ or $\PZ$ boson mass, respectively.
Additional information from jet substructure is used to reduce the background from multijet production.
No evidence is found for a signal and upper limits on the resonance production cross section are set
as function of the resonance mass. The results are interpreted in the
context of the bulk graviton model, heavy vector triplet \PWpr{} and \PZpr{} resonances, and excited quark resonances $\PQq^*$.
For the heavy vector triplet model~B, we exclude at 95\% confidence level spin-1 resonances with degenerate masses below 3.8\TeV and singlet \PWpr and \PZpr resonances with masses below 3.2 and 2.7\TeV, respectively. In the case of a singlet \PWpr resonance masses between 3.3 and 3.6\TeV can be excluded additionally.
In the narrow-width bulk graviton model, production cross sections are excluded in the range from 36.0\unit{fb} for a resonance mass of 1.3\TeV, to the most stringent limit of 0.6\unit{fb} for high resonance masses above 3.6\TeV. Exclusion limits are set at 95\% confidence level on the production of excited quark resonances $\PQq^*$ decaying to $\PQq\PW$ and $\PQq\PZ$ for masses less than 5.0 and 4.7\TeV, respectively.

\ifthenelse{\boolean{cms@external}}{\clearpage}{}

\begin{acknowledgments}

We congratulate our colleagues in the CERN accelerator departments for the excellent performance of the LHC and thank the technical and administrative staffs at CERN and at other CMS institutes for their contributions to the success of the CMS effort. In addition, we gratefully acknowledge the computing centres and personnel of the Worldwide LHC Computing Grid for delivering so effectively the computing infrastructure essential to our analyses. Finally, we acknowledge the enduring support for the construction and operation of the LHC and the CMS detector provided by the following funding agencies: BMWFW and FWF (Austria); FNRS and FWO (Belgium); CNPq, CAPES, FAPERJ, and FAPESP (Brazil); MES (Bulgaria); CERN; CAS, MoST, and NSFC (China); COLCIENCIAS (Colombia); MSES and CSF (Croatia); RPF (Cyprus); SENESCYT (Ecuador); MoER, ERC IUT, and ERDF (Estonia); Academy of Finland, MEC, and HIP (Finland); CEA and CNRS/IN2P3 (France); BMBF, DFG, and HGF (Germany); GSRT (Greece); OTKA and NIH (Hungary); DAE and DST (India); IPM (Iran); SFI (Ireland); INFN (Italy); MSIP and NRF (Republic of Korea); LAS (Lithuania); MOE and UM (Malaysia); BUAP, CINVESTAV, CONACYT, LNS, SEP, and UASLP-FAI (Mexico); MBIE (New Zealand); PAEC (Pakistan); MSHE and NSC (Poland); FCT (Portugal); JINR (Dubna); MON, RosAtom, RAS, RFBR and RAEP (Russia); MESTD (Serbia); SEIDI and CPAN (Spain); Swiss Funding Agencies (Switzerland); MST (Taipei); ThEPCenter, IPST, STAR, and NSTDA (Thailand); TUBITAK and TAEK (Turkey); NASU and SFFR (Ukraine); STFC (United Kingdom); DOE and NSF (USA). \hyphenation{Rachada-pisek}

Individuals have received support from the Marie-Curie programme and the European Research Council and EPLANET (European Union); the Leventis Foundation; the A. P. Sloan Foundation; the Alexander von Humboldt Foundation; the Belgian Federal Science Policy Office; the Fonds pour la Formation \`a la Recherche dans l'Industrie et dans l'Agriculture (FRIA-Belgium); the Agentschap voor Innovatie door Wetenschap en Technologie (IWT-Belgium); the Ministry of Education, Youth and Sports (MEYS) of the Czech Republic; the Council of Science and Industrial Research, India; the HOMING PLUS programme of the Foundation for Polish Science, cofinanced from European Union, Regional Development Fund, the Mobility Plus programme of the Ministry of Science and Higher Education, the National Science Center (Poland), contracts Harmonia 2014/14/M/ST2/00428, Opus 2014/13/B/ST2/02543, 2014/15/B/ST2/03998, and 2015/19/B/ST2/02861, Sonata-bis 2012/07/E/ST2/01406; the Thalis and Aristeia programmes cofinanced by EU-ESF and the Greek NSRF; the National Priorities Research Program by Qatar National Research Fund; the Programa Clar\'in-COFUND del Principado de Asturias; the Rachadapisek Sompot Fund for Postdoctoral Fellowship, Chulalongkorn University and the Chulalongkorn Academic into Its 2nd Century Project Advancement Project (Thailand); and the Welch Foundation, contract C-1845.

\end{acknowledgments}
\bibliography{auto_generated}

\cleardoublepage \appendix\section{The CMS Collaboration \label{app:collab}}\begin{sloppypar}\hyphenpenalty=5000\widowpenalty=500\clubpenalty=5000\textbf{Yerevan Physics Institute,  Yerevan,  Armenia}\\*[0pt]
A.M.~Sirunyan, A.~Tumasyan
\vskip\cmsinstskip
\textbf{Institut f\"{u}r Hochenergiephysik,  Wien,  Austria}\\*[0pt]
W.~Adam, F.~Ambrogi, E.~Asilar, T.~Bergauer, J.~Brandstetter, E.~Brondolin, M.~Dragicevic, J.~Er\"{o}, M.~Flechl, M.~Friedl, R.~Fr\"{u}hwirth\cmsAuthorMark{1}, V.M.~Ghete, J.~Grossmann, J.~Hrubec, M.~Jeitler\cmsAuthorMark{1}, A.~K\"{o}nig, N.~Krammer, I.~Kr\"{a}tschmer, D.~Liko, T.~Madlener, I.~Mikulec, E.~Pree, D.~Rabady, N.~Rad, H.~Rohringer, J.~Schieck\cmsAuthorMark{1}, R.~Sch\"{o}fbeck, M.~Spanring, D.~Spitzbart, W.~Waltenberger, J.~Wittmann, C.-E.~Wulz\cmsAuthorMark{1}, M.~Zarucki
% ...
\vskip\cmsinstskip
% ...
1:~~Also at Vienna University of Technology, Vienna, Austria\\
% ...

\end{sloppypar}
\end{document}